\documentclass{nature}

\usepackage{wrapfig}
\usepackage{color} 
\usepackage{multirow}
\usepackage{graphicx}
\usepackage{hyperref}
\usepackage{academicons}
\usepackage{xcolor}
\usepackage{array}
\usepackage{url}
\usepackage{float}
\newfloat{extFig}{ht}{pdf}
\floatname{extFig}{Extended Data Fig.}

\usepackage{orcidlink}

\makeatletter
\let\saved@includegraphics\includegraphics
\AtBeginDocument{\let\includegraphics\saved@includegraphics}
\renewenvironment*{figure}{\@float{figure}}{\end@float}
\makeatother

\def\amin{\ifmmode^{\prime}\else$^{\prime}$\fi}
\def\asec{\ifmmode^{\prime\prime}\else$^{\prime\prime}$\fi}

\def\simgt{\lower.5ex\hbox{$\; \buildrel > \over \sim \;$}}
\def\simlt{\lower.5ex\hbox{$\; \buildrel < \over \sim \;$}}

\newcommand\chandra{\textit{Chandra}}

\newcommand\xmm{\textit{XMM-Newton}}

\newcommand\juno{{\it Juno}}

\newcommand\nustar{\textit{NuSTAR\/}}

\newcommand{\apj}{Astrophys. J.}
\newcommand{\apjs}{Astrophys. J. Suppl. Ser.}
\newcommand{\apjl}{Astrophys. J. Lett.}
\newcommand{\nat}{Nature}
\newcommand{\aap}{Astron. Astrophys.}

\title{Observation and origin of non-thermal hard X-rays from Jupiter}

\author{Kaya Mori$^{1}$ \orcidlink{0000-0002-9709-5389}, Charles Hailey$^{1}$ \orcidlink{0000-0002-3681-145X}, Gabriel Bridges$^{1}$ \orcidlink{0000-0002-6653-4975}, Shifra Mandel$^{1}$ \orcidlink{0000-0002-6126-7409}, Amani Garvin$^{1}$, Brian Grefenstette$^{2}$ \orcidlink{0000-0002-1984-2932}, William Dunn$^{3,4,5}$ \orcidlink{0000-0002-0383-6917}, Benjamin J. Hord$^{6,7}$ \orcidlink{0000-0001-5084-4269}, Graziella Branduardi-Raymont$^{3,4}$ \orcidlink{0000-0002-6620-6357}, John Clarke$^{8}$, Caitriona Jackman$^{9}$ \orcidlink{0000-0003-0635-7361}, Melania Nynka$^{8}$ \orcidlink{0000-0002-3310-1946} and Licia Ray$^{10}$ \orcidlink{0000-0003-3727-602X}}

\begin{document}

\maketitle
\begin{affiliations}
 \item Columbia Astrophysics Laboratory, Columbia University, New York, NY 10027, USA
 \item Cahill Center for Astronomy and Astrophysics, California Institute of Technology, Pasadena, CA 91125, USA
 \item Mullard Space Science Laboratory, Department of Space and Climate Physics, University College London, Dorking, UK
 \item The Centre for Planetary Science at UCL/Birkbeck, London, UK
 \item Harvard-Smithsonian Center for Astrophysics, Smithsonian Astrophysical Observatory, Cambridge, MA 02138, USA
 \item Department of Astronomy, University of Maryland, College Park, MD 20742-2421, USA
 \item NASA Goddard Space Flight Center, Greenbelt, MD 20771, USA
 \item Kavli Institute For Astrophysics and Space Research, Massachusetts Institute of Technology, Cambridge, MA 02139, USA
 \item Astronomy and Astrophysics Section, Dublin Institute for Advanced Studies, Dublin, Ireland
\item Department of Physics, Lancaster University, Lancaster, UK
\end{affiliations}

\newpage
\begin{abstract}
Electrons accelerated on Earth by a rich variety of wave scattering or stochastic processes\cite{Li2014,Lorentzen2000} generate hard non-thermal X-ray bremsstrahlung up to $\simgt 1$~MeV\cite{Millan2002,Foat1998} and power Earth's various types of aurorae. Although Jupiter's magnetic field is an order of magnitude larger than Earth's, space-based telescopes have previously detected X-rays only up to $\sim 7$ keV\cite{BR2007}. On the basis of theoretical models of the Jovian auroral X-ray production\cite{Barbosa1990, Waite1992, Singhal1992}, X-ray emission in the $\sim2$--$7$~keV band has been interpreted as thermal (arising from electrons characterized by a Maxwell-Boltzmann distribution) bremsstrahlung\cite{BR2007, Wibisono2020}. Here we report the observation of hard X-rays in the 8--20 keV band from the Jovian aurorae, obtained with the NuSTAR X-ray observatory. The X-rays fit to a flat power-law model with slope $0.60\pm0.22$  -- a spectral signature of non-thermal, hard X-ray bremsstrahlung. We determine the electron flux and spectral shape in the kiloelectronvolt to megaelectronvolt energy range using coeval in situ measurements by the Juno spacecraft’s JADE and JEDI instruments. Jovian electron spectra of the form we observe have previously been interpreted to arise in stochastic acceleration, rather than coherent acceleration by electric fields\cite{Mauk2017}. We reproduce the X-ray spectral shape and approximate flux observed by NuSTAR, and explain the non-detection of hard X-rays by Ulysses\cite{Hurley1993}, by simulating the non-thermal population of electrons undergoing precipitating electron energy loss, secondary electron generation and bremsstrahlung emission in a model Jovian atmosphere.  The results highlight the similarities between the processes generating hard X-ray auroras on Earth and Jupiter, which may be occurring on Saturn, too.
\end{abstract}

NuSTAR, launched in 2012, is the first space-based focusing hard X-ray telescope\cite{Harrison2013}. Given its broad  energy band (3--79~keV) and sub-arcminute angular resolution, NuSTAR provides sufficient sensitivity to detect hard X-ray emission from Jupiter and spatially resolve the northern and southern aurorae. From 2015 to 2018, we conducted five NuSTAR observations of Jupiter (with a  total exposure time of 600 ks), four of which coincided with perijoves (PJ) 6, 7 and 12 and apojove (AJ) 7 of Juno's orbits (Extended Data Table~\ref{tab:obs}). The NuSTAR observations during PJ7 and AJ7 overlapped  with XMM-Newton observations, which are included in the subsequent analysis. In each of the observations, NuSTAR detected X-ray emission from Jupiter with net count rates of $\sim1.2\times10^{-3}$ counts per second on average, and significance ranging between $3$ and $5\sigma$ in the 3--20~keV band. We found no significant detection ($\simgt2\sigma$) above 20 keV. Detection significance is defined as $S = \frac{N_{\rm T} - N_{\rm B}}{\sqrt{N_{\rm T} + N_{\rm B}}}$, where $N_{\rm T}$ and $N_{\rm B}$ are total (source+background) and background counts, extracted from a circle with $r=45$\asec\ (which contains the entire planet) and an annulus at $r=60$--75\asec, respectively. As we did not detect significant flux variability among the NuSTAR observations (Methods), we combined imaging and spectral data from all NuSTAR observations, allowing us to explore the persistent hard X-ray emission above 8~keV with improved photon statistics. Unlike the recent timing studies with XMM-Newton and Chandra\cite{Dunn2016, Jackman2018}, any variable components related to the solar wind are likely to be time averaged in our NuSTAR analysis. A recent XMM-Newton observation revealed that the persistent auroral X-ray emission is largely due to charged particles accelerated inward from the plasma disk of Io\cite{Wibisono2020}, which is the Jovian reservoir of electrons and ions emitted from the volcanically active moon Io\cite{Cowley2001}.

\begin{figure}[h!]
\begin{center} 
\includegraphics[width=9cm]{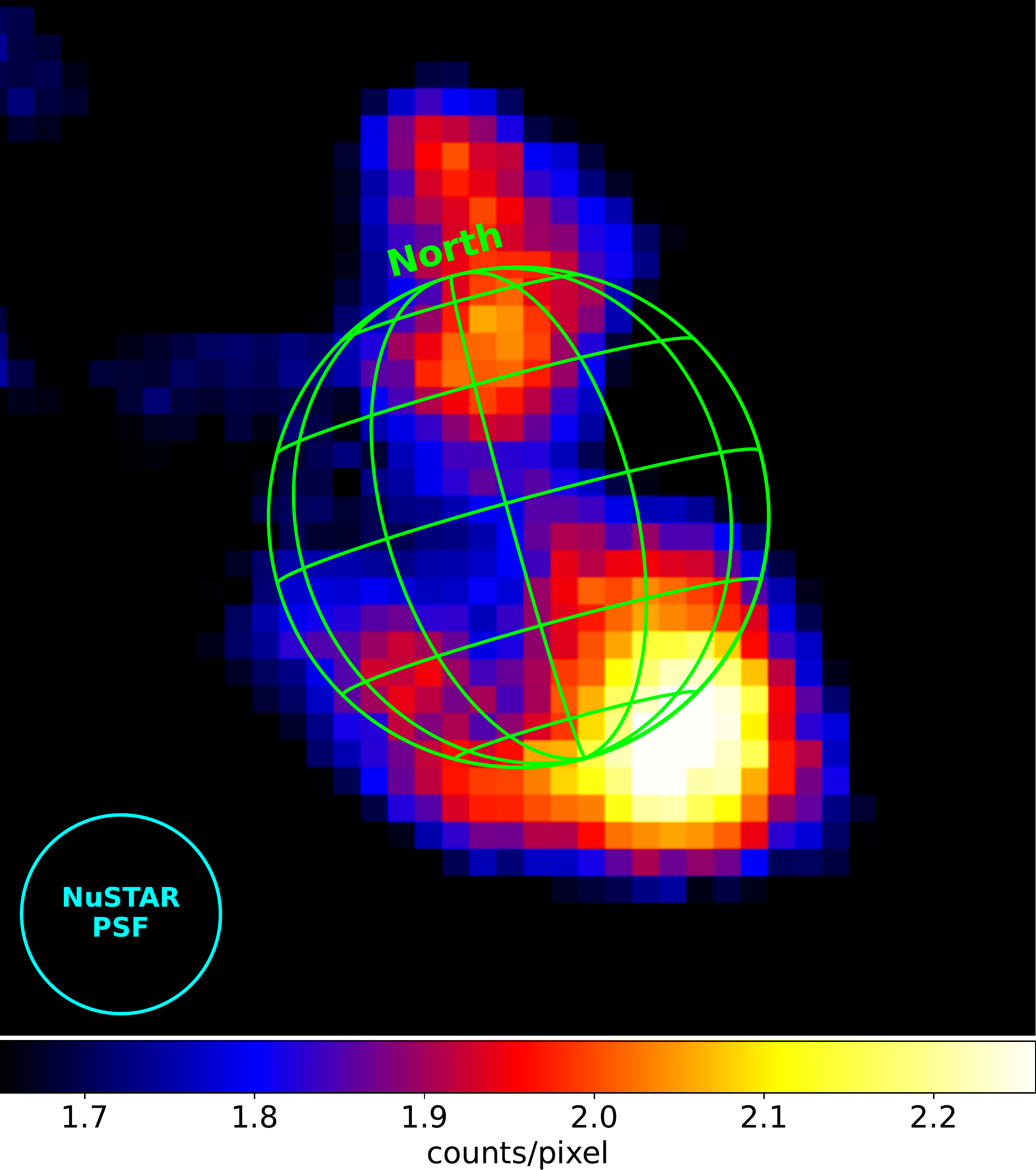}
\end{center} 
\caption{NuSTAR 8--20~keV image of Jupiter overlaid on a graticule. The north pole is indicated by 'North' on the graticule, which shows the geometry of the planet during PJ12 ($r= 42.6$\asec).. The angular size of Jupiter varies between $r=37$\asec\ and 45\asec\ during the five NuSTAR observations. We combined the two focal plan module (FPMs) images and smoothed by a gaussian kernel with $\sigma$ = 6 pixels (15\asec). The NuSTAR PSF (full-width at half-maximum of 18\asec\ , shown in the lower left corner) is comparable to the size of each auroral region, thus it is not feasible to correlate the hard X-ray emission with the UV oval or soft X-ray polar region. The northern and southern auroral regions, defined by a $r=25$\asec\ circle around each pole, yielded 114 and 229 net counts, respectively, in the 8--20 keV band. } 

\label{fig:nustar_image} 
\end{figure}

We detected hard X-ray emission from Jupiter in the 8--20~keV band with $7\sigma$ significance in the combined images. Figure~\ref{fig:nustar_image} shows a NuSTAR 8--20~keV image of Jupiter in the planet's co-moving frame, clearly resolving hard X-ray emission from the two auroral regions. Above $\sim20$ keV, we found no significant detection exceeding $\sim2\sigma$. The NuSTAR observations show higher fluxes from the southern aurora, by a factor of $2.1\pm0.1$ and $2.0\pm0.2$ in the 3--20 and 8--20~keV bands, respectively. The observed divergence in brightness is intrinsic and not due to the $7\%$ visibility difference between the auroral regions (Methods). The brighter southern aurora above 3~keV is in stark contrast to the softer X-ray band, where ion line emission is predominant and the northern aurora is consistently brighter\cite{Dunn2017}. We also confirmed that the northern aurora is brighter than its southern counterpart by a factor of $2.9\pm0.1$ in the $0.3\mathrm{-}3$~keV EPIC images of the XMM-Newton observations, which were taken simultaneously with two of the NuSTAR observations (AJ7 and PJ7). This contrast between the aurorae is further emphasized by the often incoherent X-ray pulsation signals\cite{Dunn2017}. Together, these observations suggest that the X-ray emission of the southern aurora may be more controlled by the energetic flow of electrons, and the northern aurora by precipitation of sulfur and oxygen ions and charge exchange processes, since the southern pole exhibits more persistent and higher electron currents, as revealed by recent Juno/MAG observations\cite{Kotsuaros2019}.

\begin{figure}[h!]
\begin{center} 
\includegraphics[width=16cm]{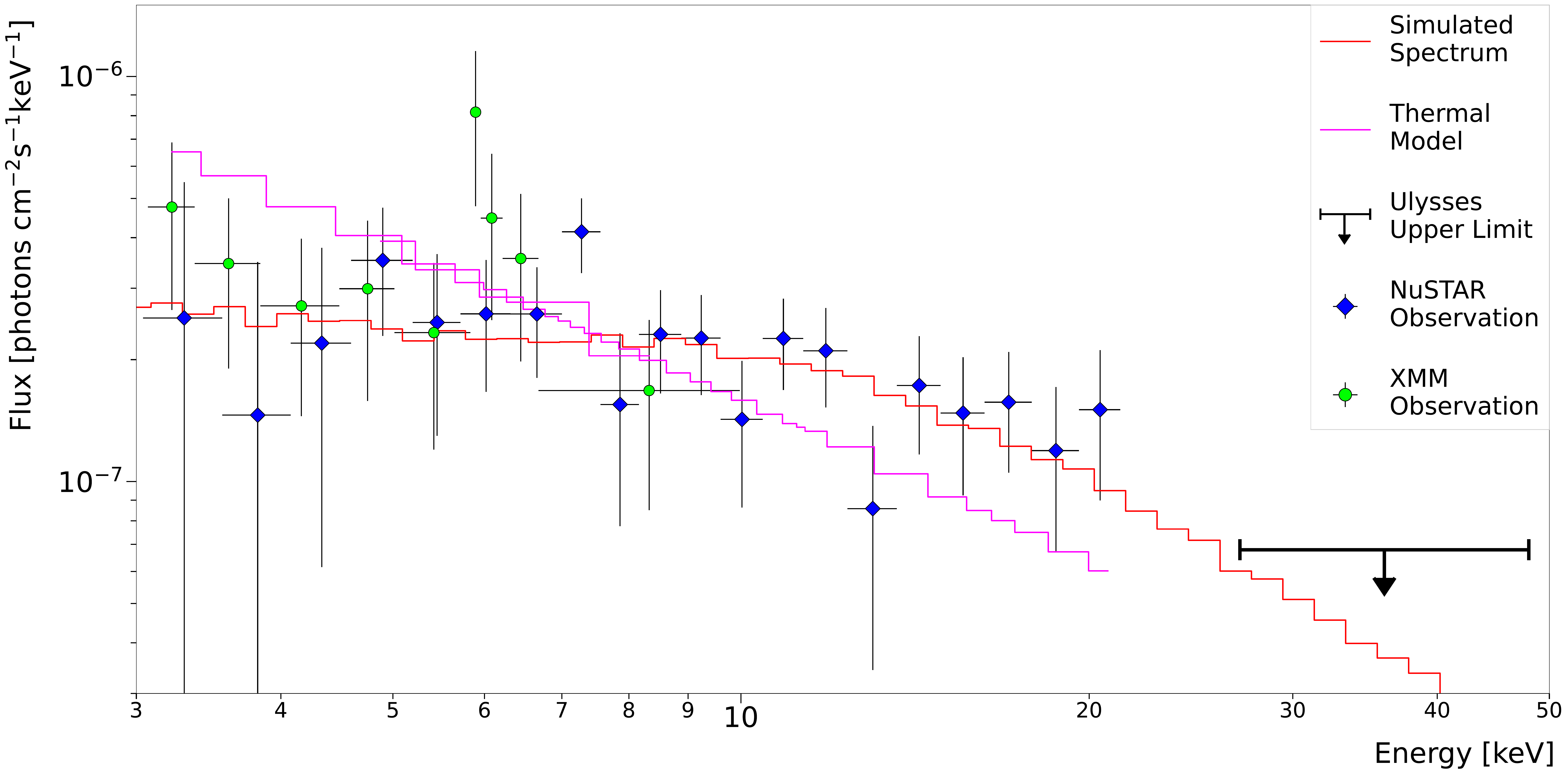}
\end{center} 
\caption{XMM-Newton-EPIC and NuSTAR flux spectra of Jupiter with the simulated spectrum and best-fit thermal bremsstrahlung model.
Both NuSTAR and EPIC spectra are binned to a minimum detection significance of $\sigma \geq2$. The NuSTAR spectra were rebinned in the plot for better visualization. The horizontal and vertical error bars represent the energy bin sizes and 1--$\sigma$ statistical errors, respectively. The 27--48~keV flux upper limits (3--$\sigma$) obtained by the in-situ Ulysses measurements\cite{Hurley1993} are indicated by the black arrow. The fit with the simulated spectrum ($\chi^2_\nu = 1.1$ for 35 d.f.) is better than that of the thermal bremsstrahlung model ($\chi^2_\nu = 1.5$ for 34 d.f.). The Akaike information criterion test yields $\Delta\rm{AIC} = 14.9$, indicating that the simulation model is preferred over the thermal bremsstrahlung model by a relative likelihood of $3\times10^{6}$. The simulated spectrum's flux normalization has been multiplied by a factor of 3.6 in order to match the X-ray flux observed by XMM-Newton and NuSTAR. }
\label{fig:xray_spectra} 
\end{figure}

We found no significant difference in the spectral hardness ratio between the northern and southern aurorae (Methods). This allowed us to perform joint spectral analysis for the two emission regions, improving fit statistics. We extracted NuSTAR and XMM-Newton-EPIC spectra from a $r=45$\asec\ circle around the Jovian center. 
NuSTAR and EPIC background spectra were taken from an annular region at $r=60\mathrm{-}75$\asec and a nearby source-free rectangular region, respectively. Joint NuSTAR and EPIC spectra are shown in Figure~\ref{fig:xray_spectra}. The NuSTAR and EPIC spectra were fitted to a power-law model with the best-fit photon index $\Gamma = 0.60\pm0.22$ (with a reduced chi-squared $\chi^2_\nu = 1.09$ for 34 degrees of freedom (d.f.)). In contrast, high-temperature thermal bremsstrahlung ($kT\sim30$--$200$~keV), previously suggested by XMM-Newton observations\cite{Wibisono2020}, has a markedly different shape and yields a poor fit in the 3--20 keV band ($\chi^2_\nu = 1.5$ for 34 d.f.). Our simulation of a thermal bremsstrahlung model, using telescope response files, resulted in a much softer X-ray spectrum with $\Gamma > 1.3$. The observed hard spectral index is a signature of non-thermal bremsstrahlung emission, and is only discernible from thermal emission through fitting the  broadband NuSTAR data.

\begin{figure}[h!]
\begin{center} 
\includegraphics[width=17cm]{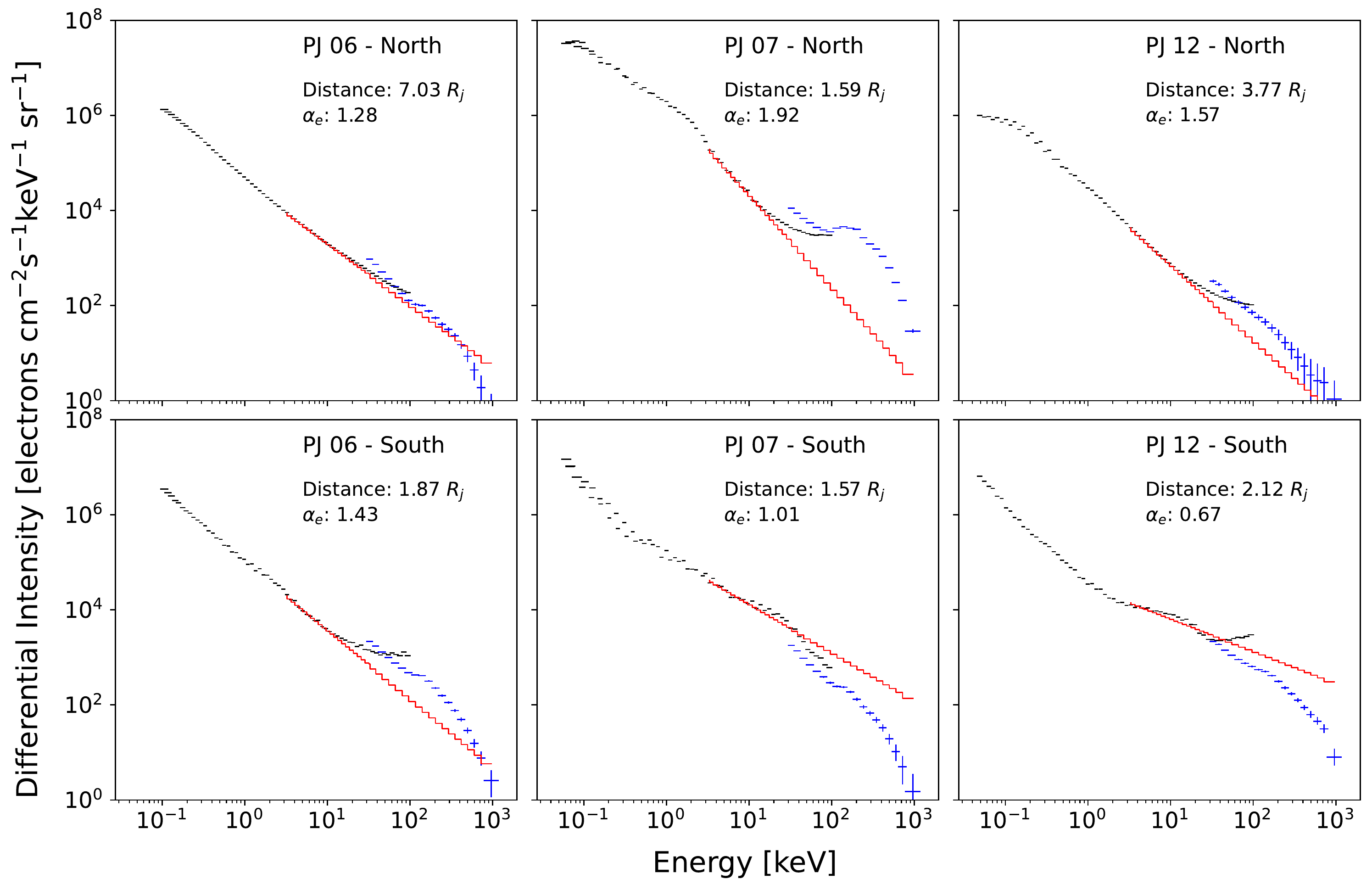}
\end{center} 
\caption{JADE and JEDI energy spectra of  precipitating electrons for PJ6, PJ7, and PJ12. The JADE (black) and JEDI (blue) energy spectra of precipitating electrons in the north (top panels) and south (bottom panels) for PJ6 (left), PJ7 (middle), and PJ12 (right). The power-law models fitted from 3 keV to 1~MeV, corresponding to the energy range of electrons that produce 3--20 keV bremsstrahlung X-rays in the Jovian atmosphere, are overlaid in red. The best-fit $\alpha_e$ and the distance of Juno from the Jovian center (in unit of the Jovian radius $R_{J}$) are listed in each panel. The corresponding Juno observations and more details are listed in Extended Data Table~\ref{tab:juno}.}  
\label{fig:electron_spectra} 
\end{figure}

The simultaneous Juno and NuSTAR data presents a unique opportunity to exploit observed, non-thermal electron fluxes and spectra, and an appropriate model, to make ab initio predictions of the hard X-ray flux and spectrum for direct comparison with the X-ray observations. Consequently we simulated X-ray bremsstrahlung spectra from precipitating electrons measured by Juno's JADE and JEDI instruments. We extracted JADE and JEDI electron spectra from PJ6, PJ7 and PJ12, ranging in energy from 0.1 keV to 1 MeV\cite{Clark2017}. The downward electron current into the atmosphere was selected by confining the range of pitch  angles at $\theta \le 44^\circ$ (northern aurora) and $\theta \le 37^\circ$ (southern aurora)\cite{ALLEGRINI2017} with respect to the magnetic field orientation measured by the Juno/MAG instrument\cite{Connerney2018}(e.g. See Extended Data Figure \ref{fig:juno_trajectory} for Juno's magnetic footprint in PJ12). Broadband JADE/JEDI electron spectra from $E_e \sim 1$~keV to $\sim1$~MeV are well characterized by a power-law model ($N (E_e) \sim E_e^{-\alpha_e}$)  with $\alpha_e = $  0.7--1.9 (Extended Data Table \ref{tab:juno} and Figure~\ref{fig:electron_spectra}). We interpret these near-smooth spectra, with no evidence of a peak in phase space, as indicative of a stochastic or broadband (as opposed to coherent) acceleration process, as has been argued for spectra from previous Juno observations\cite{Clark2018}. We found that the electron spectral shape, solely extracted from the Juno data, is insensitive to the selection of a loss cone range. We then simulated X-ray bremsstrahlung spectra using the GEANT4 particle propagation simulator\cite{Agostinelli2003}. In each simulation, we injected a beam of 100 million electrons -- which were generated randomly from a power-law model with mean spectral index $\alpha_e = 1.3$, calculated from the observed JADE/JEDI electron spectra (Figure~\ref{fig:electron_spectra}) -- into a stratified spherical shell that mimicked the Jovian atmosphere (Methods). The most accurately determined chemical composition and density/temperature profile of the atmosphere in the polar regions have been implemented in our simulation\cite{ATREYA2003,Atreya1981}. All primary electrons, electrons generated by collisional ionization and X-ray bremsstrahlung photons are tracked. We recorded X-ray photons that escape the model atmosphere to construct bremsstrahlung X-ray spectra.  

\begin{figure}[h!]
\begin{center} 
\includegraphics[width=15cm]{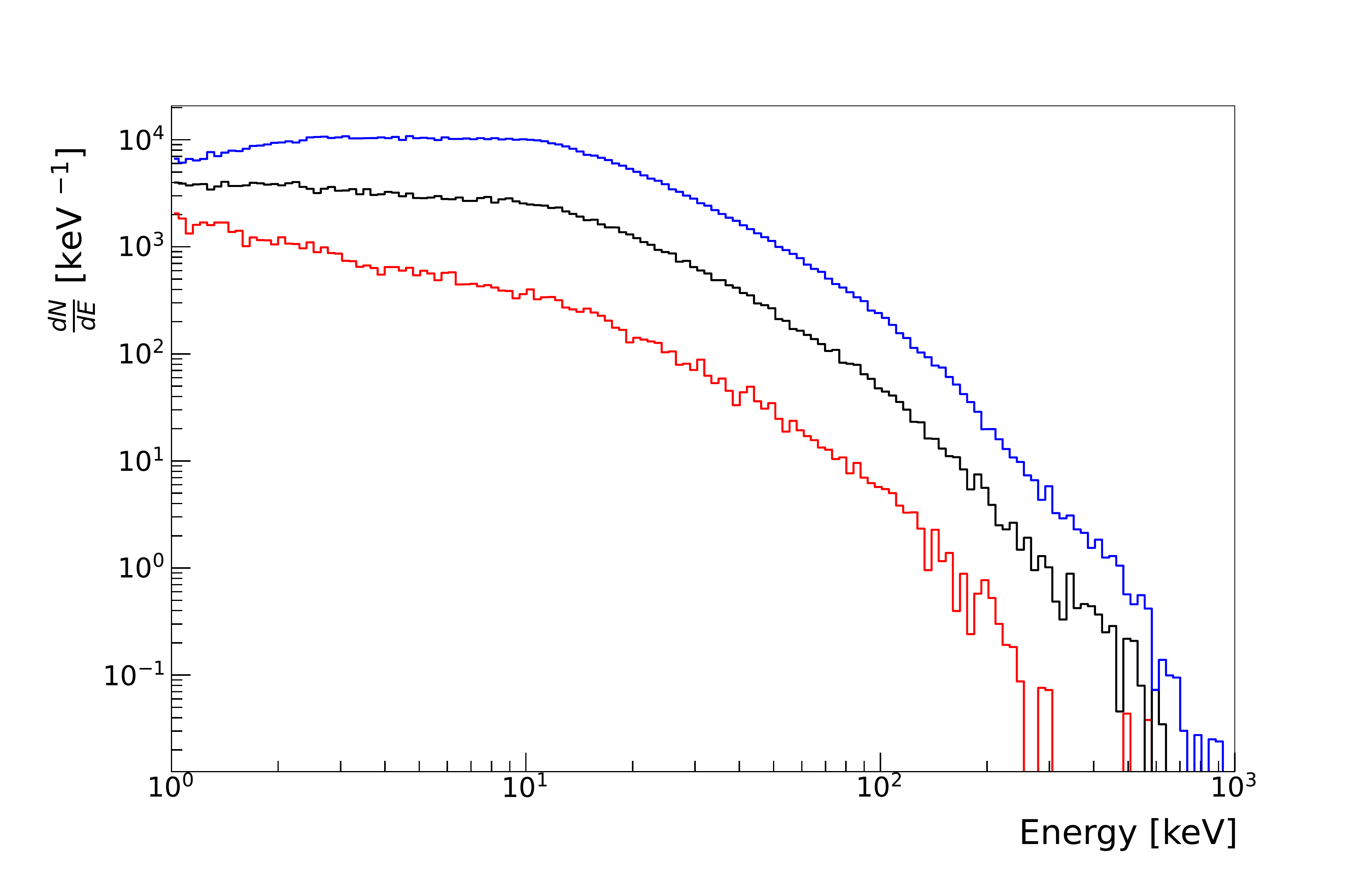}
\end{center} 
\caption{Simulated spectra of X-ray photons escaping from the model atmosphere for three different electron spectra. We input the mean electron spectrum (black) from the Juno observations listed in Extended Data Table~\ref{tab:juno} into the GEANT4 simulator, as well as the hardest (blue) and softest (red) spectra observed by Juno. The spectra are plotted in counts per energy bin versus photon energy. The simulated X-ray spectra are characterized by a flat power-law component ($\Gamma \sim$ 0.03--0.4) up to $E\sim10$~keV, followed by a softer power-law component ($\Gamma \sim 1$). Fitting a single power-law model to the simulated X-ray spectra in the 3--20 keV band, where we obtained EPIC and NuSTAR spectra, yields $\Gamma = 0.2$--0.7, which is consistent with the observed X-ray photon index ($\Gamma = 0.60 \pm 0.22$).}   
\label{fig:xray_model_spectrum} 
\end{figure}

Inserting the full JADE and JEDI electron spectrum into the GEANT4 simulation that realistically models the Jovian atmosphere yielded a spectrum that reproduces the features observed in the broadband NuSTAR and XMM-Newton X-ray data. These include both the observed X-ray power law with an effective $\Gamma \approx 0.6$ (3--20~keV) and a gradual spectral softening at $\simgt20$~keV. The observed and simulated spectral softening explain the non-detection of X-rays above 27 keV by Ulysses. This consistency confirms that the observed hard X-rays are indeed generated by these electrons. One of the earlier XMM-Newton observations, fitting the narrow 2--7~keV energy band, also detected a flat power-law component and an extrapolated flux at 27--48~keV that was inconsistent with the Ulysses non-detection, indicating a rollover below 50 keV.\cite{BR2007} These key features are robust; they are apparent over the entire range of electron spectral indices ($\alpha_e = 0.7$--1.9) measured by Juno (Figure~\ref{fig:xray_model_spectrum}). The predicted X-ray flux, calculated by combining all the downward electron data from the PJ6, PJ7 and PJ12 orbits, is lower than the observed flux by a factor of 1.4--4, depending on the electron spectrum selection (Methods). This constitutes excellent agreement, considering the sensitivity of the result to the spatial and temporal variability of the electron flux, statistical uncertainties associated with the selection of the electron loss cone, the precise size of the X-ray emitting area, and the omission of magnetic mirroring. Our simulations indicate that the bulk of the bremsstrahlung X-rays originate at an altitude $\sim200$~km (corresponding to a neutral hydrogen column density $N_{\rm H} \sim 10^{22}$~cm$^{-2}$) above Jupiter’s surface (defined as a pressure of 1 bar). This is substantially deeper in the stratosphere than findings from previous work, which assumed that the X-rays were generated by thermal bremsstrahlung, resulting in peak emission at altitudes of $\sim260$--$340$~km\cite{Singhal1992}. The comparable upwelling flux of electrons observed by Juno was ignored in our simulations. Those electrons must originate at much higher altitudes, where X-ray emission is negligible due to the low atmospheric density. Otherwise, the upwelling electron spectrum would have been substantially different than what Juno observed, due to electron energy losses in the denser stratosphere.

Our interpretation of Jovian X-ray spectra relied on forward folding a detailed model of the Jovian atmosphere with the relevant X-ray emission physics, using coeval Juno electron data as initial input, as has been done for similar Earth X-ray auroral analysis\cite{Woodger2015}. But the X-ray spectra alone encode substantial information about the most energetic electrons. More than 95\% of the NuSTAR band X-rays arise from electrons with $\simgt100$~keV energies. Unfolding Jovian hard X-ray spectra can this provide the novel capability of extracting correlated electron spectra when such information is not otherwise available. This high energy electron probe is possible because the electron-energy dependence of the X-ray bremsstrahlung cross-section and the higher molecular densities required for X-ray emission both favor more energetic electrons. Although electron spectra from coherent magnetic-field-aligned acceleration differ from those produced by stochastic acceleration, similar considerations suggest that unfolding hard X-ray spectra is unlikely to permit discrimination between the acceleration processes; the X-ray observation times would have to be much longer, as the periods of such coherent electron acceleration are rarer than the stochastic processes\cite{Mauk2017}. 

Broadband NuSTAR X-ray observations have provided clear evidence of non-thermal, hard X-ray bremsstrahlung from a Solar System body other than Earth. The observed spectral shape and flux arise naturally from non-thermal electron precipitation,  energy loss and subsequent X-ray emission in the Jovian atmosphere. The coeval JADE/JEDI electron spectra associated with our X-ray spectra show the unambiguous signatures of stochastic or broadband acceleration of non-thermal electrons, in analogy to the interpretation established in previous work by the JEDI/JADE teams\cite{Mauk2017}, and similar to processes responsible for the diffuse aurorae on Earth\cite{Millan2002,Foat1998,Li2014,Lorentzen2000}. Unrealistically high electron temperatures of $kT \simgt100$~keV\cite{Wibisono2020}, such as are typically seen in stellar objects, are no longer required to fit the harder part of the X-ray spectrum. And the non-thermal X-ray bremsstrahlung spectrum naturally softens above 20~keV, explaining the Ulysses non-detection in the 27--48~keV band\cite{Hurley1993}. Further hard X-ray bremsstrahlung studies of Jupiter will provide a fruitful approach to elucidating Jovian magnetospheric physics, as such studies have on Earth. Moreover, an intense, bi-directional, non-thermal electron spectrum has also been observed from the magnetosphere of Saturn\cite{Saur2006}, and holds out the prospect that deep, hard X-ray observations may yet reveal a third planet with non-thermal hard X-ray bremsstrahlung.

\begin{methods}

\subsection{Introduction}

X-ray emission is ubiquitous in the Solar System. Among the bodies detected in X-rays (aside from the Sun) are the terrestrial planets (Venus, Earth, Mars); three of the gas giants (Jupiter, Saturn, Uranus); the Moon and the Jovian moons Europa, Io and Ganymede; comets; and the rings of Saturn\cite{Bhardwaj2007,Dunn2020}. While a wide variety of emission mechanisms is associated with these objects, one of the most common astrophysical emission mechanisms – bremsstrahlung, in which (de)accelerated electrons generate X-rays – has only been observed on the Earth and Jupiter\cite{Bhardwaj2006,Imhof1974}.
Earth’s magnetosphere (the rotating magnetic field – plasma system) contains a thermal electron population (characterized by a Maxwell-Boltzmann distribution) that generates soft X-rays when electrons are coherently accelerated by powerful electric fields aligned with magnetic field lines connecting to high geomagnetic latitudes. In contrast, harder X-rays from a few keV to $\simgt 1$~MeV are generated through its relativistic electron populations. 
Extensive balloon and satellite observations over decades have elucidated a rich variety of theoretical mechanisms for precipitating electrons with non-thermal spectra that can generate non-thermal, hard X-ray bremsstrahlung. Such mechanisms include stochastic pitch angle scattering of electrons out of the radiation belts via whistler mode chorus and electromagnetic ion cyclotron (EMIC) waves\cite{Li2014,Lorentzen2000}, and 
Alfvenic-wave electron interactions in the magnetosphere
\cite{Li2014,Lorentzen2000,Chaston2008}. 

Ganymede, Jupiter and the trans-Jovian planets, like Earth, all have active dynamos supporting a magnetosphere and aurorae\cite{Christensen2019,Bhardwaj2000,Feldman2000}.  But Jupiter, whose magnetic field ($\sim12$~Gauss) is stronger than that of Ganymede and these other planets by a factor of ~100 and ~10 respectively, has extremely powerful aurorae. In contrast to Earth and its electron-ion population magnetically trapped in the radiation belts, the Jovian reservoir of electrons and ions is provided by its moon Io. Io is located at $\sim6R_{\rm J}$, deep in the Jovian magnetosphere, and is volcanically active due to the enormous Jovian tidal forces; it ejects a large outflow of particles, mostly composed of SO$_2$ molecules, at a rate of approximately 1 metric ton per second\cite{Delamere2004}. The neutral particles are ionized, with $\simgt50$\% remaining in the system as plasma\cite{Dougherty2017}. These Iogenic particles migrate outwards due to Jupiter’s centrifugal force and form a plasma disk in the equatorial plane at $R \sim (15$ -- $40) R_{\rm J}$ -- the Io plasma disk\cite{Cowley2001}.  The Io plasma disk is the ultimate reservoir of charged particles whose acceleration inward along magnetic field lines connected to the polar regions produces Jupiter’s intense auroral emission from the radio to X-ray bands.

Since the discovery of UV and X-ray auroral emission from Jupiter in 1979\cite{Broadfoot1979,Metzger1983}, a series of \textit{HST} and \chandra\ observations have dissected the emission regions and mechanisms in great detail. The UV emission originates from oval regions around the poles as a result of $\sim10$--$500$~keV electrons exciting or ionizing H$_2$ molecules in the upper atmosphere\cite{Gerard2019}, while X-ray emission is more concentrated in the polar regions\cite{Gladstone2002}.  Subsequent X-ray observations with \xmm\ detected two distinct components (at less and greater than $\sim2$~keV, respectively) in the auroral X-ray emission. The \simlt2~keV emission completely dominates the X-ray flux and has been attributed to charge exchange (CX) line emission from highly stripped sulfur and oxygen ions interacting with atmospheric hydrogen molecules\cite{BR2004}.  
A more recent \xmm\ observation in June 2017, through fitting the CX emission lines with various ion species, including carbon, oxygen and sulfur, revealed that the Iogenic component, as opposed to the solar wind ions, plays the major role in the soft X-ray emission\cite{Wibisono2020}. But a thermal bremsstrahlung component was still required to adequately fit the $\sim2$--$7$~keV data. Because of the narrow energy band of  \xmm\, the electron temperature was poorly constrained ($kT \sim100$--$300$~keV)\cite{Wibisono2020}. 
Earlier XMM-Newton observations in 2003 were equally well-fit with either poorly constrained thermal bremsstrahlung or a power-law model\cite{BR2007}. Above the XMM-Newton energy band, in-situ measurements by the Ulysses spacecraft yielded no detection of hard X-rays in the 27-48 keV band\cite{Hurley1993}. Despite the narrow ~2-7 keV energy band, one of the earlier XMM-Newton observations detected a flat power-law component indicating a rollover below ~50 keV\cite{BR2007}.    

Barbosa (1990)\cite{Barbosa1990} was the first to propose, in analogy to Earth’s coherent acceleration of thermal electrons by geomagnetically-aligned electric fields, that such primary electrons precipitate in the Jovian atmosphere, continuously slow down, and generate bremsstrahlung X-rays. Both Waite et al. (1992)\cite{Waite1992} and Singhal et al. (1992)\cite{Singhal1992} also adopted this approach by assuming a Maxwellian electron beam at the top of the atmosphere. For instance, Singhal et al. self-consistently explained the entire soft X-ray flux and spectrum observed by \textit{Einstein} below 2 keV and the electron densities measured by \textit{Voyager 2}, assuming a characteristic primary electron temperature of $kT \sim30$~keV. 
This is much higher than Earth's thermal electron population, with $kT \sim$ few--$10$~keV.  The nature of the continuum X-ray emission takes on new urgency in light of recent results from the Juno and Galileo missions\cite{Paranicas2018,Kollmann2018}. These missions detected a non-thermal electron energy distribution extending up to $\sim$~MeV energies. And more surprisingly, the electron spectra measured by the JADE and JEDI instruments on \juno\ revealed that the dominant electron acceleration process in Jovian aurorae is not coherent, B-field aligned electric fields, but rather stochastic or broadband processes associated with wave scattering or acceleration\cite{Mauk2017}. Such (non-thermal) processes produce the most energetic X-ray bremsstrahlung on Earth -- much higher than $\sim7$~keV. The newer Jovian observations thus prompt a reexamination of the nature of the $\sim2$--$7$~keV tail of the Jovian X-ray spectra, and to search for higher energy X-ray emission.

\subsection{X-ray observations and data reduction }
NuSTAR observed Jupiter five times from 2015 to 2018 with a total exposure of $\sim600~$ksec. Besides the first observation in 2015, the NuSTAR observations coincided with Juno's perijoves (PJ6, PJ7 and PJ12) or apojove (AJ7) as shown in Extended  Data Table~\ref{tab:obs}. In each NuSTAR observation, a series of multiple NuSTAR pointings were consecutively performed by tracking the (moving) target close to the on-axis position. Using the Jet Propulsion Laboratory's HORIZONS ephemeris data, we corrected all photon event positions to Jupiter's co-moving frame. The absolute astrometric accuracy of NuSTAR is $\pm5$\asec\ (90\%)\cite{Harrison2013} which was confirmed by comparing the optical and NuSSTAR positions of a background AGN (SDSS J092412.11$+$161135.5)\cite{Abazajian2009} in the 2015 observation. For the other NuSTAR observations in 2017--2018, no X-ray sources  were observed in the field of view. All NuSTAR data were processed using the nupipeline command\cite{Harrison2013}. We processed EPIC data from the two XMM-Newton observations coinciding with the NuSTAR observations in 18 June and 10 July 2017 using SAS version 16.1.0. Similar to the NuSTAR data, we modified all photon event positions to Jupiter's comoving frame. 

\renewcommand{\tablename}{Extended Data Table}

\begin{table}
\small  
\caption{NuSTAR observations of Jupiter}
\centering
\begin{tabular}{c c c c c c c}
\hline\hline
Observation & \hspace{1.6pt} Distance \hspace{1.6pt} & \hspace{1.6pt} Exposure \hspace{1.6pt} & \hspace{1.6pt} Net$^{\rm a}$ \hspace{1.6pt} & \hspace{1.6pt} Detection \hspace{1.6pt} & \hspace{1.6pt} Juno \hspace{1.6pt} & Simultaneous \\[-8pt]
date & [AU] & [ksec] & counts & $\sigma$ & orbit & observations \\
\hline
01/30/2015 & 4.4 & 102.6 & 105 & 3.3 & --- & --- \\
05/16/2017 & 4.7 & 134.5 & 132 & 3.2  & PJ6 & \chandra\  \\
06/18/2017 & 5.1 & 101.5 & 130 & 3.4 & AJ7 & \chandra, {\it XMM}, {\it HST}  \\
07/10/2017  & 5.4 & 134.2 & 163 & 4.2 & PJ7 & \chandra, {\it XMM}, {\it HST}  \\
04/01/2018 & 4.6 & 126.3 & 197 & 5.0 & PJ12 & \chandra \\
\hline
\end{tabular}

$^{\rm a}$ Net count rates in the 3--20 keV band. Both FPMA and FPMB counts are combined. Source and background counts were extracted from a $r=45$\asec\ circle and a $r=60\mathrm{-}75$\asec\ annular region around the Jovian center, respectively. 
\label{tab:obs}
\end{table}
 
\subsection{Visibility of the Jovian auroral regions }
To compute the visibility of the northern and southern aurora, the left handed Sys III latitudes and longitudes for the main auroral ovals were obtained from the LASP Magnetospheres of Outer Planets Group (\url{https://lasp.colorado.edu/home/mop/missions/juno/trajectory-information}). The Jupiter ephemeris was gathered from JPL’s HORIZONS database. The auroral ovals were orthographically projected and the total visible projected area was computed for every interval within a \nustar\ observation. The areas of the northern and southern aurorae were compared in this way and we found that the northern aurora was on average $7.2\%$ more visible than the southern aurora during the NuSTAR observations.

\subsection{Background analysis } 
The NuSTAR background below $\sim30$~keV is dominated by stray-light photons -- X-ray photons from point sources and diffuse X-ray emission outside the field of view and therefore not reflected by the optics. The Galactic latitude of Jupiter's position during the NuSTAR observations ranged from 32$^\circ$ to 59$^\circ$. Stray-light background from the Galactic Ridge X-ray emission did not substantially affect our analysis; this was confirmed by the lack of Fe emission lines in the background spectra. Based on the nuskybkg tool\cite{Wik2014}, we found that the NuSTAR background is not spatially uniform on the detector plane, mostly due to the stray-light background from cosmic X-ray diffuse emission. Therefore, we extracted NuSTAR background counts and spectra from an annulus close to the source, at $r=60$--75\asec\ around the Jupiter position. XMM-Newton background spectra were extracted from a rectangular region near the source, avoiding the detector chip gaps.

\subsection{Variability analysis } 
We studied long-term X-ray variation using NuSTAR count rates between the different NuSTAR observations. To minimize statistical errors, we generated light curves for each observation after combining the FPMA and FPMB data, using a circular extraction region of $r=45$\asec. Extended Data Table~\ref{tab:obs} shows the source counts and detection significance for each NuSTAR observation. The detection significance is defined as $S = \frac{N_{\rm T} - N_{\rm B}}{\sqrt{N_{\rm T} + N_{\rm B}}}$, where $N_{\rm T}$ and $N_{\rm B}$ are total (source+background) and background counts, respectively. 
We did not find any significant variation in the 3--20~keV lightcurves between the NuSTAR observations. There has been no prominent solar activity during the NuSTAR observations. Our search for short-term (hours) variability signals using the Bayesian Block algorithm\cite{Scargle2013} yielded no detection, possibly due to the limited statistics. Due to the lack of variability, we stacked NuSTAR images and spectra in the subsequent analysis.   

\subsection{Spectral analysis }
We extracted NuSTAR source spectra and generated detector response files as well as effective area files using the nuproducts command\cite{Harrison2013}. We performed spatially-resolved spectral analysis using hardness ratios and stacked NuSTAR spectra from each auroral region (defined as a $r=25$\asec\ circle around the centroid of X-ray spot at each pole). We obtained $114$ and $229$ net counts (8--20~keV band) from the northern and southern regions, respectively, by combining all the NuSTAR observations. We defined the hardness ratios as $HR = (H-S)/(H+S)$, where $H$ and $S$ are the net counts in 3--8 keV and 8--20~keV, respectively. The hardness ratios between the two aurorae, $HR = 0.15\pm0.24$ (north) and $0.10\pm0.11$ (south), do not differ statistically significantly. In order to improve the statistics, we stacked 3--30~keV NuSTAR spectra extracted from a $r=45$\asec\ circle around the Jovian center (including both aurorae) and from all the observations listed in Extended Data Table~\ref{tab:obs}. Background spectra were extracted from an annular region at $r=60$--75\asec\ around the Jovian center. NuSTAR spectra were adaptively rebinned so that each energy bin contains enough source counts to ensure more than $2\sigma$ significance over the background. To fit NuSTAR spectra, we adopted the 3--22~keV band, above which the background dominates. EPIC source photons were extracted from a circle of $r=45$\asec\ around the Jovian center, and background photons from a nearby source-free rectangular region. We adopted a lower energy bound for EPIC data of 3~keV in order to ensure that there is no contamination from the CX line emission \simlt2~keV. Jointly fitting 3--10~keV EPIC and 3--30~keV NuSTAR spectra with a power-law model, using the detector response and telescope effective area files in XSPEC, yielded a photon index ($\Gamma = 0.60\pm0.22$) with $\chi^2_\nu = 1.02$ for 34 d.f. Fitting a thermal bremsstrahlung model resulted in the maximum temperature of $kT = 200$~keV allowed in the XSPEC model; the temperature is poorly constrained and the fit yields an unacceptably large reduced chi-squared value of $\chi^2_\nu = 1.51$ for 34 d.f. In Figure \ref{fig:xray_spectra}, we present the unfolded NuSTAR and EPIC flux spectra, as well as the Ulysses flux upper limit in the 27--48 keV band.

\subsection{Juno JADE and JEDI data analysis }
We extracted in-situ electron data from Juno collected during the three perijoves (PJ6, PJ7, and PJ12) coinciding with the NuSTAR observations. The JADE\cite{JADE} and JEDI\cite{JEDI} instruments aboard Juno measure the energies and pitch angles of ions and electrons in the 0.1--100~keV and 30--1000~keV band, respectively. The JADE and JEDI level 3 files were gathered from the Planetary Plasma Interactions node of the NASA Planetary Data System as well as the magnetometer data from the MAG instrument (in payload coordinates)\cite{JADEDATA, JEDIDATA, MAGDATA}. SPICE kernels were also acquired from the NAIF data node of the PDS\cite{SPICE}. JEDI-E level 3 files contain differential intensity and pitch angle data of electrons, whereas JADE-E level 3 files provide count rates, energies, and look directions of electrons in despun spacecraft coordinates. 
We converted the JADE-E electron data to differential intensity distribution (in electrons cm$^{-2}$ s$^{-1}$ keV$^{-1}$ sr$^{-1}$), following the methodology of Allegrini et al.\cite{Allegrini2020}. A conversion from despun spacecraft coordinates to pitch angles was performed using the transformation matrix data provided in the level 3 files and the SPICE kernel data, as well as the MAG magnetic field data. We converted the channel number to kinetic energy (keV) for each JEDI electron event using the calibration files. For each of the three perijoves (PJ6, PJ7 and PJ12), we selected two time intervals corresponding to Juno's passages over the northern and southern poles (Extended Data Table~\ref{tab:juno}). For each of the six time intervals, electron event times were selected when Juno's magnetic footprint was close to the auroral ovals\cite{AURORALOVALS} ($1.5 - 2.6 R_{\rm J}$ or 42,500 - 117,500 km above the surface) and JADE-E recorded high net count rates($\sim10^8$ cts\,s$^{-1}$ above the background level). 

Only precipitating electrons are considered in this investigation. As shown in Extended Data Figure \ref{fig:depth}, the X-ray emission peaks deep in the stratosphere, where the equivalent hydrogen column density is $N_{\rm H} \sim 10^{22}$~cm$^{-2}$. Simulations using GEANT demonstrate that the upwelling electron spectrum would be substantially modified by ionization energy losses from that observed by Juno unless electrons originate at much higher altitudes than the stratosphere, and at these higher altitudes the particle densities are too low for efficient production of bremsstrahlung. We segmented the differential intensity data into downward and upward electrons using loss cone intervals (defined for each pole)\cite{ALLEGRINI2017}. We found that the spectral shape of the precipitating electrons was insensitive to the selection of a loss cone range, even between pitch angles $\theta = 12^{\circ}$ and $\theta = 90^{\circ}$. Pitch angles are locally defined with respect to the magnetic field orientation measured by the Juno/MAG instrument. The lower limit of the pitch angle range, corresponding to the integral range of differential electron flux, was set to the field of view of a single JEDI detector. Note that the MAG data are well characterized by the JRM09 model which represents the global magnetic field geometry\cite{Connerney2018}, while the perturbation by the Birkeland current is negligible ($< 1\%$)\cite{Kotsuaros2019}. Thus, the pitch angle data obtained by the JADE, JEDI and MAG instruments accurately reflect the electron current's directions above the auroral regions. Ultimately, we adopted the values found in Allegrini et al.\cite{ALLEGRINI2017} ($\theta < 44^{\circ}$ for the northern aurora and $\theta < 37^{\circ}$ for the southern aurora) to construct the flux data for precipitating electrons. The electron flux normalization varies linearly with the loss cone angle range. JADE is composed of three separate instruments that, when combined, offer complete pitch angle coverage. However, one of these instruments (JADE 300) malfunctioned in flight and had been switched off, resulting in incomplete pitch angle coverage. To rectify this effect, we excluded electron data for the time intervals when the full loss cone coverage was not obtained by JADE. 

We produced spectra of the precipitating electrons for each pole and JADE/JEDI instrument by averaging over the period for which Juno's magnetic footprint overlapped the auroral oval and across all look directions within the defined loss cones. 
Merging the observations of these two instruments into a single electron spectrum required some modification in the  30--100 keV band where the two instruments overlap. Since JADE has higher energy resolution than JEDI in this energy range, we rebinned JADE spectra to match up with the JEDI energy bins. This re-binned spectra was then averaged with the JEDI data for fitting purposes. The resulting electron spectra, constructed from the JADE and JEDI data, spanned from 0.1 keV to 1 MeV with only a small discrepancy between the two instruments. Figure~\ref{fig:electron_spectra} shows the precipitating electron spectra of the northern and southern aurorae, taken by JADE and JEDI instruments during PJ6, PJ7 and PJ12. 
We fit a power-law model ($N(E_e) \propto E_e^{-\alpha_e}$) to characterize JADE, JEDI and joint spectra (Extended Data Table~\ref{tab:juno}). In several instances, a power-law model did not yield a good fit to the joint JADE + JEDI spectra (for example, the north pole orbit in PJ7). Otherwise, the JADE, JEDI and joint spectra are well characterized by a power-law model yielding the best-fit spectral indices ranging from $\alpha_e = 0.7$ to $1.9$ (Extended Data Table~\ref{tab:juno}). The power-law form of the electron spectra of Figure~\ref{fig:electron_spectra}, with the possible exception of north pole PJ7, is evidence of electrons accelerated by stochastic or broadband acceleration\cite{Mauk2017}. The north pole PJ7 spectrum shows a positive slope just above $\sim100$~keV, but the spectral slope $\alpha_e \sim-0.65$ to $-0.95$ is shallower than the $\alpha_e \simlt-1$ characteristic of coherent acceleration by magnetically aligned electric fields. Consequently we calculated the average electron spectral index, including all six Juno orbits ($\bar\alpha_e = 1.30$). For the reason noted just above, the average spectral index is unchanged by the exclusion of north pole PJ7 data.
Note that we did not weight the spectral index by the measured electron fluxes since the Juno's orbit altitude varied between the observations (thus likely leading to different electron fluxes). Hereafter, we assumed that the precipitating electron spectra follow a power-law model of $\alpha_e = 1.3$. 

\begin{table}
\small
\caption{Juno orbits coincident with NuSTAR observations and electron spectral parameters}
\centering
\begin{tabular}{c c c c c c c}
\hline\hline
\emph{Juno} & \hspace{14pt} Orbit \hspace{14pt} & \hspace{14pt} Pole \hspace{14pt} & \hspace{14pt} Time \hspace{14pt} & \hspace{14pt} $R_J$ \hspace{14pt} & $\alpha_e$ & Electron flux \\[-10pt]
Orbit &  Date &  &  &  &  & [electrons/cm$^{2}$/s]\\
\hline
\multirow{2}{*}{PJ 6} 
    & \multirow{2}{*}{05/19/2017} & North & 00:57:11 & 7.03 & $1.26 \pm .05$ & $1.16\times 10^{5}$\\
                            && South & 06:49:00 & 1.87 & $1.41 \pm .61$ & $3.50\times 10^{5}$\\ 
\multirow{2}{*}{PJ7} 
    & \multirow{2}{*}{07/11/2017} & North & 01:19:00 & 1.59 & $1.89 \pm .14$ & $2.79\times 10^{6}$\\
                            && South & 02:30:00 & 1.57 & $1.01 \pm 0.03$ & $5.78\times 10^{5}$\\     
\multirow{2}{*}{PJ 12} 
    & \multirow{2}{*}{04/01/2018} & North & 07:32:00 & 3.77 & $1.54 \pm 0.08$ & $4.82\times 10^{4}$\\
                            && South & 10:42:30 & 2.12 & $0.67 \pm 0.01$ & $5.36\times 10^{5}$\\ \hline  

\end{tabular}
\\Note: $\alpha_e$ is the best-fit spectral index of JADE + JEDI electron spectra in each Juno passage. 
The last column lists the downward electron flux in the 3keV -- 1 MeV band. 
\label{tab:juno} 
\end{table}

\subsection{Spatial and temporal variability of the JADE/JEDI electron flux } 
As the cumulative observation time of the Juno orbits (7 ksec) is a small fraction of the NuSTAR exposure, the spatial and temporal variability of the precipitating electron flux may affect the flux normalization of our model X-ray spectra. In order to investigate the electron flux variability, we extracted JADE/JEDI electron data from a 2--minute time interval of all available Juno orbits (PJ1--PJ24 excluding PJ2) around their perijove points. The 2--minute interval was selected to ensure that Juno would complete two rotations on either side of the perijove point. The total electron flux above 3  keV, after averaging over all time intervals and look directions, was calculated for each orbit. We found a factor of 2.1 variability between different Juno orbits, representing the temporal variability over weeks to months. In addition, as shown in Extended Data Fig.~\ref{fig:juno_trajectory}, the electron count rate changed drastically in each Juno orbit. Thus, the selection of a loss cone and X-ray emission area (through the pitch angle range and time interval from which electron data are extracted), both of which cannot be  completely determined from the current in-site and X-ray observations, can result in additional flux uncertainties. These effects likely caused the electron flux variability by a factor of 1.9--3.0 and 1.2--1.4 at the northern and southern pole, respectively, between PJ6, PJ7 and PJ12 (Extended Data Table~\ref{tab:juno}). Overall, we estimated the electron (and thus X-ray) flux errors up to a factor of $\sim 2-3$.  

\begin{extFig}[h!]
\begin{center} 
\includegraphics[width=15cm]{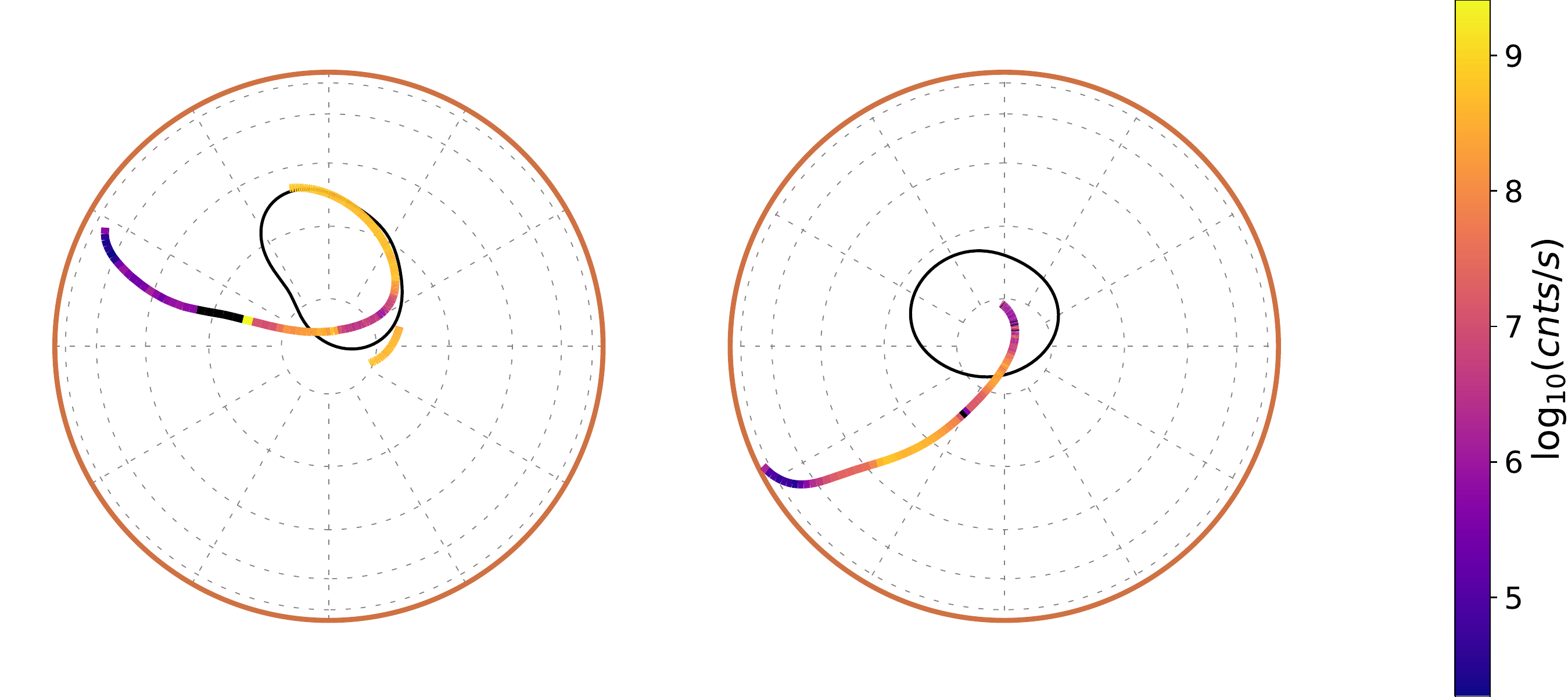}
\end{center} 
\caption{Magnetic footprint of Juno as viewed from above the north (left) and south (right) poles in PJ12. Black ovals represent the UV ovals\cite{AURORALOVALS}. The color bar indicates the net electron counts obtained by JADE across all energy channels and look directions (in logarithmic scale). A region in the Juno orbit (indicated by black), where background counts exceeded signal counts, was removed from our analysis. We obtained coordinates for the auroral ovals and JRM09 magnetic footprint\cite{Connerney2018} from the MOP LASP website. All coordinates are SYS III.}   
\label{fig:juno_trajectory} 
\end{extFig} 

\subsection{Simulation of electron bremsstrahlung spectra in the Jovian atmosphere } 
Given the complexity of electron propagation and bremsstrahlung X-ray emission in the Jovian atmosphere, we utilized the widely-used particle propagation simulator GEANT4\cite{Agostinelli2003,Allison2006}. Our Monte-Carlo simulation approach, by taking into account the electron energy distribution measured in-situ, the most updated particle/atomic database and a more realistic Jovian atmosphere profile, supercedes the original calculation of Singhal et al. in 1992\cite{Singhal1992} and is similar to the approach utilized for Earth X-ray modeling by Woodger et al. in 2015\cite{Woodger2015}. Our simulation employs the Electromagnetic Penelope Physics List, which agrees most accurately with experimental bremsstrahlung results below 3 MeV\cite{PANDOLA2015}. In our simulation, we modeled a stratified Jovian atmosphere based on the density/temperature profiles of the polar regions obtained by Atreya and colleagues\cite{Atreya1981,ATREYA2003} and Livengood et al.\cite{Livengood1990}. The model atmosphere is a spherical shell with the Jovian radius and a thickness of 80,000~km which  corresponds to Juno's perijove altitude, divided into four distinct regions (see below) and composed of 80 total layers of molecular/atomic hydrogen, helium atoms, and methane. The fraction of other species is negligible. 
Each layer's thickness was set such that none of the parameters (temperature, density, or pressure) varies by more than a factor of 2  across the layer. We found this configuration optimal for achieving the most robust results within reasonable CPU time, after testing  various cases. Since initial simulations demonstrated that X-rays were exclusively produced below 700 km of altitude, we decreased the thickness of these layers to 10 km and achieved a smooth X-ray emissivity curve as a function of altitude (Extended  Data  Fig.~\ref{fig:depth}). The finer altitude depths removed numerical glitches which are artificially produced as a result of the typical bremsstrahlung interaction length being shorter than the grid size. 

The model atmosphere consists of four regions, each of which was characterized by attributes in its profile. (1) Region 1: The lowest of these regions consists of 43 layers between an altitude of $h=0--430$ km measured from the Jovian surface at $1~ R_{\rm J}$. It is characterized by a narrow temperature range of 150--200~K but a steep density variation. These layers are composed entirely of molecular hydrogen, helium, and methane mixed at the average fractional ratios for Jupiter's atmosphere\cite{ATREYA2003}. (2) Region 2: Above this was a region at $h=430\mathrm{-}1400$ km characterized by a rapid change in temperature (a gradient of $\sim1$~K/km) connecting the low temperature region below it to the high temperature ($\sim$1200 K) region above it\cite{Atreya1981}. Within this region, the chemical composition varies dramatically, largely due to the steep temperature gradient: methane quantities were negligible in this region and helium becomes increasingly sparse until the atmosphere is dominated entirely by molecular hydrogen. Then, atomic hydrogen begins to become more prominent as hydrogen molecules become dissociated at higher temperatures. (3) Region 3: The top atmospheric region at $h=1400\mathrm{-}5000$~km is characterized by an  exponential density profile. It is assumed that this region has a constant temperature of 1,200~K, but our simulation is insensitive to this specific temperature. (4) Region 4: The Juno observations during PJ6, Pj7, and PJ12 occurred at an average height of 83,800~km, far above the atmospheric regions mentioned above. To fill in the gap, we introduced a single homogeneous layer of 75,000~km height and composed of atomic hydrogen. In this region we assumed a constant density of 50~cm$^{-3}$, similar to the ion density value measured by \juno\cite{Valek2019}. Given the small total hydrogen column density ($N_{\rm{H}} < 10^{13}$~cm$^{-2}$), this above-the-atmosphere region has a negligible impact on electron deceleration and X-ray emission/absorption. As mentioned above, the average JADE and JEDI electron spectra are well characterized by a single power-law model with $\alpha_e \approx 1.3$ up to $\sim1$~MeV. In each simulation, we randomly injected 100 million primary electrons between 3 keV and 1 MeV from a power-law distribution with $\alpha_e = 1.3$. While a majority of the primary electrons are eventually absorbed in the atmosphere, the number of injected electrons is sufficient to characterize X-ray spectra and compare well with the NuSTAR and EPIC spectra. We collected X-ray photons escaping from the spherical atmosphere at its top. Most of the precipitating electrons emit bremsstrahlung X-rays at the upper atmospheric depths corresponding to $N_{\rm H} \simlt 10^{22}$~cm$^{-2}$ (mostly in Region 1 and 2). As shown in Extended Data Fig.~\ref{fig:depth}, X-ray emissivity drops sharply below $\sim200$ km in Region 1 where the neutral hydrogen column density dramatically increases ($N_{\rm H} \simgt 10^{25}$~cm$^{-2}$). On the other hand, X-rays are scarce above $\sim600$ [km] since the hydrogen density is not high enough to emit bremsstrahlung X-rays. Note that X-ray photons should escape from the auroral regions at 80--90$^\circ$ degrees with respect to the pole (in the direction of Earth) in order to be observed by X-ray telescopes\cite{Ozak2010}. However, X-ray bremsstrahlung photons above $E\sim3$ keV do not suffer from limb darkening since photo-absorption and Compton scattering are insignificant at the atmospheric depths where they are emitted. 

\begin{extFig}[h!]
\begin{center} 
\includegraphics[width=15cm]{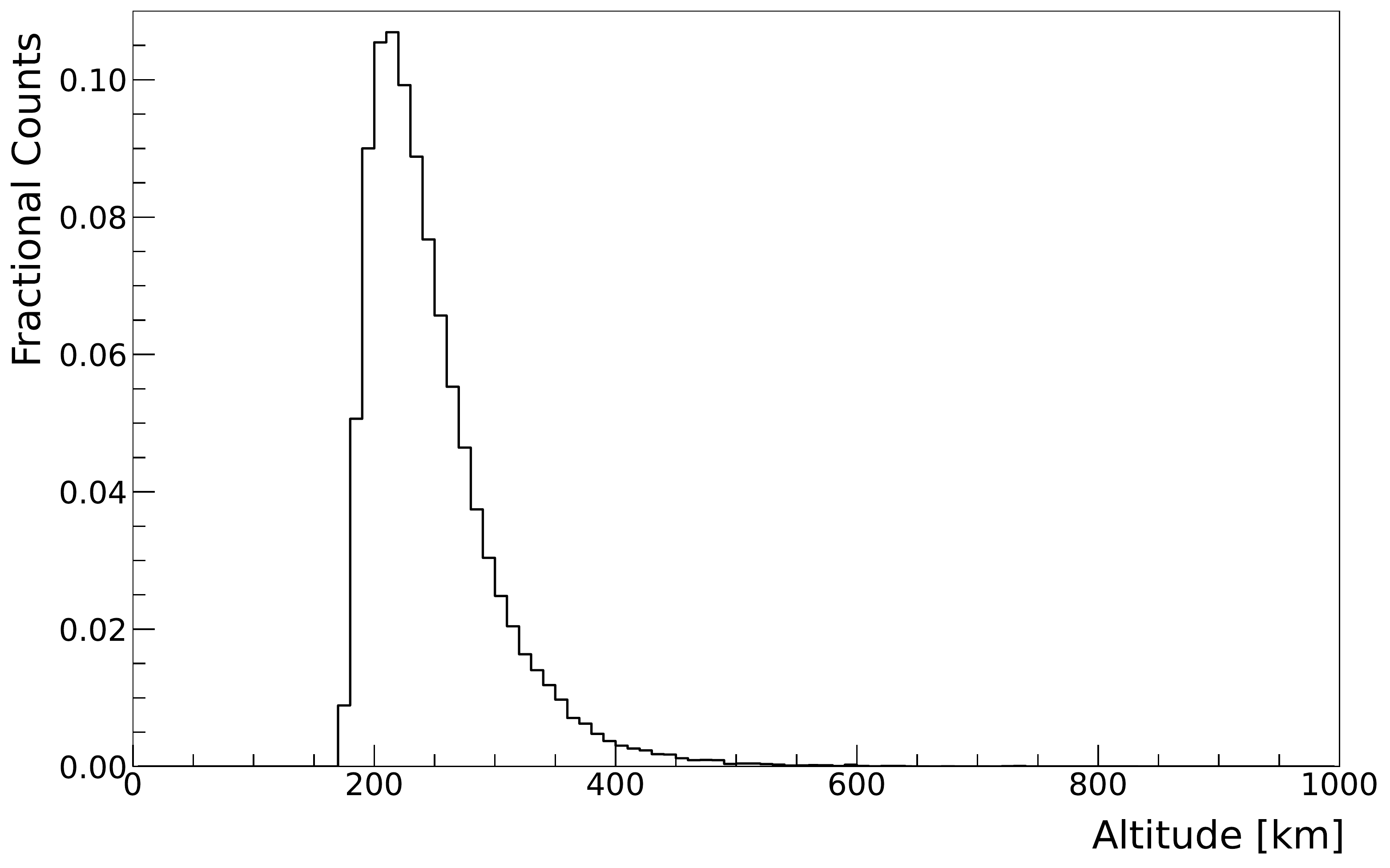}
\end{center} 
\caption{Counts of escaping X-rays as a function of the altitude [km] (which is measured from 1 R$_j$). The counts drop sharply below $h\sim200$ [km] as X-rays are heavily absorbed. The altitude range shown in the plot cover Region 1 and 2 described in the Method. Note that most of escaping X-rays are from Region 1 ($h < 430$~km). }   
\label{fig:depth} 
\end{extFig} 

To compute the total X-ray luminosity emitted from both auroral regions, we multiplied the integrated electron flux obtained by JADE/JEDI, by an electron-to-X-ray conversion factor, and the projection of the emitting area of the two auroral regions ($A$) onto Juno's orbital altitude. For simplicity, we adopted $A = 1\times10^{20}$~cm$^2$, corresponding to the area of the UV ovals which overlap with the X-ray continuum emission above $E \sim 2$~keV\cite{BR2008}, assuming that the electron  flux is spatially uniform over the auroral regions. Since Juno's orbits during the perijoves covered only a portion of the magnetic field lines above the aurorae, this assumption may lead to errors in the predicted X-ray flux if the electron current is not spatially uniform over the oval region, since this causes electron flux variability as discussed in the previous section. From the GEANT4 simulation, we determined the ratio of the number of X-rays to the number of primary electrons in the 3--20 keV  band. This ratio is about  $5.8\times10^{-4}$. Since our simulation results showed that the bremsstrahlung X-ray emission is nearly isotropic, we computed X-ray flux by dividing the X-ray luminosity by $4\pi d^2$ where $d$ is the average distance of Jupiter ($7.24\times10^{13}$ cm; weighted by the exposure time of each NuSTAR observation) during the NuSTAR observations. 

If we adopt the average electron spectra from PJ6, PJ7 and PJ12, the estimated 3--20~keV flux of $9.1\times10^{-7}$~photons/cm$^2$/s is a factor of 3.6 smaller than the measured X-ray flux of $3.3^{+0.2}_{-0.8}\times10^{-6}$~photons/cm$^2$/s, while the observed X-ray spectral shape obtained from NuSTAR, XMM-Newton and Ulysses data are well reproduced  (Figure \ref{fig:xray_spectra}). The flux normalization discrepancy is likely due to the spatial/temporal electron flux variation over the auroral regions (as discussed in the previous section) as well as  uncertainties associated with the selection of a loss cone and X-ray emitting area. We found that the electron flux varied between different Juno perijoves by a factor of 2.1 (1-$\sigma$ standard deviation between all available perijoves) to 3 (the maximum flux variation between PJ6, PJ7 and PJ12). For instance, if we take the highest  electron fluxes measured in each pole during PJ 6, 7 and 12, the estimated 3--20~keV flux of $2.4\times10^{-6}$~photons/cm$^2$/s is consistent with the measured X-ray flux (i.e. a factor of 1.4 lower but within the statistical errors). In addition, any uncertainty associated with the loss cone and X-ray emitting area should affect the predicted X-ray flux linearly. We also took into account the possibility that multi-megaelectronvolt energy electrons populate the Jovian magnetosphere beyond the JADE/JEDI energy band ($>1$~MeV) as suggested by recent Juno observations\cite{Paranicas2018}. When we extended the best-fit power-law model up to a maximum energy $E_{\rm max} = 2$~MeV and 5~MeV, we found that the flux normalization discrepancy decreases to a factor of 2.2 and 0.96, respectively. This is due to the contribution of additional X-ray emission from higher energy electrons that terminate deeper in the atmosphere. However, we found that simulated X-ray spectra with $E_{\rm max} \simgt  1$~MeV are harder than the observed X-ray spectra (e.g., $\chi^2_\nu = 1.25$ with 35 dof for $E_{\rm max } = 2$~MeV). Furthermore, our X-ray modeling with GEANT4, consistent with the work of previous authors\cite{Woodger2015,Foat1998}, ignores magnetic mirroring. This effect should produce higher downward electron fluxes above the atmosphere at $\sim1R_{\rm J}$ (thus leading to higher X-ray fluxes) compared to the fluxes measured at higher altitudes $\sim2R_{\rm J}$ corresponding to the perijoves. This would systematically increase our modeled X-ray fluxes, bringing them more into agreement with the observations.  Given the electron flux variability, statistical errors related to the X-ray flux, and these other systematics, we consider that the observed and predicted X-rays fluxes are consistent with each other.

Simulated X-ray spectra are plotted in Figure~\ref{fig:xray_model_spectrum}, as well as in Figure~\ref{fig:xray_spectra} along with the NuSTAR and EPIC spectra. 
The predicted and observed X-ray spectra are similar to each other, exhibiting a hard spectral index ($\Gamma \approx 0.6$) followed by spectral softening (which is consistent with the Ulysses X-ray flux upper limits in the 27--48~keV), after matching the flux normalization between the spectra. We found that this characteristic X-ray spectral shape is robust by inputting electron spectral indices (ranging from $\alpha_e \sim 0.8$ to 1.7, as measured by Juno) than the average value ($\alpha_e = 1.3$) in the GEANT4 simulation. We note that deceleration of electrons in the upper atmosphere is required to reproduce the flat X-ray spectra with $\Gamma \approx 0.6$ (3--20~keV); the precise spectral shape results largely from the energy-dependence of the electron stopping power ($dE_e/dx$) increasing at lower electron energy, $E_e \simlt 1$~MeV. We also found that the Jovian atmosphere density profile reproduces the observed X-ray spectral shape well; atmosphere layers with higher column densities yielded too soft model spectra, compared to the measured X-ray photon index in the 3--20~keV band, by overproducing low-energy X-rays. Thus, we emphasize that the consistency between the observed and predicted X-ray spectra is a natural consequence of propagating the Juno electron data taken during the perijoves into a realistic Jovian atmosphere model -- there is no parameter fitting besides the X-ray flux normalization. 
Our analysis and simulation, based on the simultaneous NuSTAR, XMM-Newton and Juno observations, establish that the X-ray continuum emission from the Jovian aurorae originates from precipitating non-thermal electrons in the upper atmosphere.  

\begin{addendum}
\item[Data availability] The NuSTAR and XMM-Newton data are archived at NASA's HEASARC website. The JEDI, JADE, and MAG data are available at the Planetary Data System. The magnetic footprint of JUNO is available through the LASP MOP group's website. 
\item[Code availability] NuSTAR and XMM-Newton data are analyzed by HEASOFT and SAS analysis software, respectively. Geant4 simulation tool is publicly available at https://geant4.web.cern.ch. The simulation and data reduction code is available upon request. 
 \item[Correspondence] Correspondence and requests for materials
should be addressed to Kaya Mori~(email: kaya@astro.columbia.edu).
\end{addendum}

\end{methods}

\begin{addendum}
 \item We thank Dr. Rob Wilson for his help in analyzing JADE data. We further thank the MOP group of LASP for their work on Juno's trajectory and magnetic footprint.
 We acknowledge Dr. Daniel Wik for his help in analyzing NuSTAR background data. We are grateful for the referees for making useful comments. Support for this work by K.M was provided by NASA through NuSTAR Cycle 3 Guest Observer Program grant NNH16ZDA001N. C. M. J.'s work at DIAS was supported by the Science Foundation Ireland Grant 18/FRL/6199.
 \item[Author contributions] K.M., C.H., G.B. and S.M. wrote the manuscript and made large contributions to data analysis and interpretation. G.B. performed JADE/JEDI data analysis and GEANT4 simulations. S.M. and B.J.H. analyzed NuSTAR data.  B.G. contributed to NuSTAR data analysis and interpretation. A.G. and W.D. were involved with XMM-Newton data analysis. J.C. and M.N. conducted a feasibility study of NuSTAR observations of Jupiter. G.B-R., C.J. and L.R. interpreted the analysis results and provided insights to the Jovian aurora physics. All authors contributed to discussing the results and commenting on the manuscript.  
 \item[Competing interests] The authors declare that they have no
competing financial interests.
\item[Data availability] The NuSTAR and XMM-Newton data are archived at NASA's HEASARC website. The JEDI, JADE, and MAG data are available at the Planetary Data System. The magnetic footprint of JUNO is available through the LASP MOP group's website. 
\item[Code availability] NuSTAR and XMM-Newton data are analyzed by HEASOFT and SAS analysis software, respectively. Geant4 simulation tool is publicly available at https://geant4.web.cern.ch. The simulation and data reduction code is available upon request. 
 \item[Correspondence] Correspondence and requests for materials
should be addressed to Kaya Mori~(email: kaya@astro.columbia.edu).
\end{addendum}

\pagebreak


\begin{thebibliography}{10}
\expandafter\ifx\csname url\endcsname\relax
  \def\url#1{\texttt{#1}}\fi
\expandafter\ifx\csname urlprefix\endcsname\relax\def\urlprefix{URL }\fi
\providecommand{\bibinfo}[2]{#2}
\providecommand{\eprint}[2][]{\url{#2}}


\bibitem{Li2014}
\bibinfo{author}{{Li}, Z.} \emph{et~al.}
\newblock \bibinfo{title}{{Investigation of EMIC wave scattering as the cause
  for the BARREL 17 January 2013 relativistic electron precipitation event: A
  quantitative comparison of simulation with observations}}.
\newblock \emph{\bibinfo{journal}{Geophys. Res. Lett.}} \textbf{\bibinfo{volume}{41}},
  \bibinfo{pages}{8722--8729} (\bibinfo{year}{2014}).

\bibitem{Lorentzen2000}
\bibinfo{author}{{Lorentzen}, K.~R.} \emph{et~al.}
\newblock \bibinfo{title}{{Precipitation of relativistic electrons by
  interaction with electromagnetic ion cyclotron waves}}.
\newblock \emph{\bibinfo{journal}{J. Geophys. Res.}} \textbf{\bibinfo{volume}{105}},
  \bibinfo{pages}{5381--5390} (\bibinfo{year}{2000}).

\bibitem{Millan2002}
\bibinfo{author}{{Millan}, R.~M.}, \bibinfo{author}{{Lin}, R.~P.},
  \bibinfo{author}{{Smith}, D.~M.}, \bibinfo{author}{{Lorentzen}, K.~R.} \&
  \bibinfo{author}{{McCarthy}, M.~P.}
\newblock \bibinfo{title}{{X-ray observations of MeV electron precipitation
  with a balloon-borne germanium spectrometer}}.
\newblock \emph{\bibinfo{journal}{Geophys. Res. Lett.}} \textbf{\bibinfo{volume}{29}},
  \bibinfo{pages}{2194} (\bibinfo{year}{2002}).

\bibitem{Foat1998}
\bibinfo{author}{{Foat}, J.~E.} \emph{et~al.}
\newblock \bibinfo{title}{{First detection of a terrestrial MeV X-ray burst}}.
\newblock \emph{\bibinfo{journal}{Geophys. Res. Lett.}} \textbf{\bibinfo{volume}{25}},
  \bibinfo{pages}{4109--4112} (\bibinfo{year}{1998}).
  

\bibitem{BR2007}
\bibinfo{author}{{Branduardi-Raymont}, G.} \emph{et~al.}
\newblock \bibinfo{title}{{A study of Jupiter's aurorae with XMM-Newton}}.
\newblock \emph{\bibinfo{journal}{\aap}} \textbf{\bibinfo{volume}{463}},
  \bibinfo{pages}{761--774} (\bibinfo{year}{2007}).
\newblock \eprint{astro-ph/0611562}.

\bibitem{Barbosa1990}
\bibinfo{author}{{Barbosa}, D.~D.}
\newblock \bibinfo{title}{{Bremsstrahlung X rays from Jovian auroral
  electrons}}.
\newblock \emph{\bibinfo{journal}{J. Geophys. Res.}} \textbf{\bibinfo{volume}{95}},
  \bibinfo{pages}{14969--14976} (\bibinfo{year}{1990}).

\bibitem{Waite1992}
\bibinfo{author}{{Waite}, J., J.~H.}, \bibinfo{author}{{Boice}, D.~C.},
  \bibinfo{author}{{Hurley}, K.~C.}, \bibinfo{author}{{Stern}, S.~A.} \&
  \bibinfo{author}{{Sommer}, M.}
\newblock \bibinfo{title}{{Jovian Bremsstrahlung X rays: A Ulysses
  prediction}}.
\newblock \emph{\bibinfo{journal}{Geophys. Res. Lett.}} \textbf{\bibinfo{volume}{19}},
  \bibinfo{pages}{83--86} (\bibinfo{year}{1992}).

\bibitem{Singhal1992}
\bibinfo{author}{{Singhal}, R.~P.}, \bibinfo{author}{{Chakravarty}, S.~C.},
  \bibinfo{author}{{Bhardwaj}, A.} \& \bibinfo{author}{{Prasad}, B.}
\newblock \bibinfo{title}{{Energetic electron precipitation in Jupiter's upper
  atmosphere}}.
\newblock \emph{\bibinfo{journal}{J. Geophys. Res.}} \textbf{\bibinfo{volume}{97}},
  \bibinfo{pages}{18245--18256} (\bibinfo{year}{1992}).

\bibitem{Wibisono2020}
\bibinfo{author}{{Wibisono}, A.~D.} \emph{et~al.}
\newblock \bibinfo{title}{{Temporal and Spectral Studies by XMM-Newton of
  Jupiter's X-ray Auroras During a Compression Event}}.
\newblock \emph{\bibinfo{journal}{J. Geophys. Res. Space Phys.}} \textbf{\bibinfo{volume}{125}}, \bibinfo{pages}{e27676}
  (\bibinfo{year}{2020}).

\bibitem{Mauk2017}
\bibinfo{author}{{Mauk}, B.~H.} \emph{et~al.}
\newblock \bibinfo{title}{{Discrete and broadband electron acceleration in
  Jupiter's powerful aurora}}.
\newblock \emph{\bibinfo{journal}{\nat}} \textbf{\bibinfo{volume}{549}},
  \bibinfo{pages}{66--69} (\bibinfo{year}{2017}).
 


\bibitem{Hurley1993}
\bibinfo{author}{{Hurley}, K.}, \bibinfo{author}{{Sommer}, M.} \&
  \bibinfo{author}{{Waite}, J.~H.}
\newblock \bibinfo{title}{{Upper limits to Jovian hard X radiation from the
  Ulysses gamma ray burst experiment}}.
\newblock \emph{\bibinfo{journal}{J. Geophys. Res.}} \textbf{\bibinfo{volume}{98}},
  \bibinfo{pages}{21217--21220} (\bibinfo{year}{1993}).

  
 \bibitem{Harrison2013}
\bibinfo{author}{{Harrison}, F.~A.} \emph{et~al.}
\newblock \bibinfo{title}{{The Nuclear Spectroscopic Telescope Array (NuSTAR)
  High-energy X-Ray Mission}}.
\newblock \emph{\bibinfo{journal}{\apj}} \textbf{\bibinfo{volume}{770}},
  \bibinfo{pages}{103} (\bibinfo{year}{2013}).
\newblock \eprint{1301.7307}.

\bibitem{Dunn2016}
\bibinfo{author}{{Dunn}, W.~R.} \emph{et~al.}
\newblock \bibinfo{title}{{The impact of an ICME on the Jovian X-ray aurora}}.
\newblock \emph{\bibinfo{journal}{J. Geophys. Res. Space Phys.}} \textbf{\bibinfo{volume}{121}}, \bibinfo{pages}{2274--2307}
  (\bibinfo{year}{2016}).

\bibitem{Jackman2018}
\bibinfo{author}{{Jackman}, C.~M.} \emph{et~al.}
\newblock \bibinfo{title}{{Assessing Quasi-Periodicities in Jovian X-Ray
  Emissions: Techniques and Heritage Survey}}.
\newblock \emph{\bibinfo{journal}{J. Geophys. Res. Space Phys.}} \textbf{\bibinfo{volume}{123}}, \bibinfo{pages}{9204--9221}
  (\bibinfo{year}{2018}).

\bibitem{Cowley2001}
\bibinfo{author}{{Cowley}, S.~W.~H.} \& \bibinfo{author}{{Bunce}, E.~J.}
\newblock \bibinfo{title}{{Origin of the main auroral oval in Jupiter's coupled
  magnetosphere-ionosphere system}}.
\newblock \emph{\bibinfo{journal}{Planet. Space Sci.}} \textbf{\bibinfo{volume}{49}},
  \bibinfo{pages}{1067--1088} (\bibinfo{year}{2001}).
  
\bibitem{Dunn2017}
\bibinfo{author}{{Dunn}, W.~R.} \emph{et~al.}
\newblock \bibinfo{title}{{The independent pulsations of Jupiter's northern and
  southern X-ray auroras}}.
\newblock \emph{\bibinfo{journal}{Nat. Astron.}}
  \textbf{\bibinfo{volume}{1}}, \bibinfo{pages}{758--764}
  (\bibinfo{year}{2017}).
  
\bibitem{Kotsuaros2019}
\bibinfo{author}{{Kotsiaros}, S.} \emph{et~al.}
\newblock \bibinfo{title}{{Birkeland currents in Jupiter's magnetosphere
  observed by the polar-orbiting Juno spacecraft}}.
\newblock \emph{\bibinfo{journal}{Nat. Astron.}}
  \textbf{\bibinfo{volume}{3}}, \bibinfo{pages}{904--909}
  (\bibinfo{year}{2019}).

\bibitem{Clark2017}
\bibinfo{author}{{Clark}, G.} \emph{et~al.}
\newblock \bibinfo{title}{{Energetic particle signatures of magnetic
  field-aligned potentials over Jupiter's polar regions}}.
\newblock \emph{\bibinfo{journal}{Geophys. Res. Lett.}} \textbf{\bibinfo{volume}{44}},
  \bibinfo{pages}{8703--8711} (\bibinfo{year}{2017}).

\bibitem{ALLEGRINI2017}
\bibinfo{author}{{Allegrini}, F.} \emph{et~al.}
\newblock \bibinfo{title}{{Electron beams and loss cones in the auroral regions
  of Jupiter}}.
\newblock \emph{\bibinfo{journal}{Geophys. Res. Lett.}}
  \textbf{\bibinfo{volume}{44}}, \bibinfo{pages}{7131--7139}.

\bibitem{Connerney2018}
\bibinfo{author}{{Connerney}, J.~E.~P.} \emph{et~al.}
\newblock \bibinfo{title}{{A New Model of Jupiter's Magnetic Field From Juno's
  First Nine Orbits}}.
\newblock \emph{\bibinfo{journal}{Geophys. Res. Lett.}} \textbf{\bibinfo{volume}{45}},
  \bibinfo{pages}{2590--2596} (\bibinfo{year}{2018}).

\bibitem{Clark2018}
\bibinfo{author}{{Clark}, G.} \emph{et~al.}
\newblock \bibinfo{title}{{Precipitating Electron Energy Flux and
  Characteristic Energies in Jupiter's Main Auroral Region as Measured by
  Juno/JEDI}}.
\newblock \emph{\bibinfo{journal}{J. Geophys. Res. Space Phys.}} \textbf{\bibinfo{volume}{123}}, \bibinfo{pages}{7554--7567}
  (\bibinfo{year}{2018}).

\bibitem{Agostinelli2003}
\bibinfo{author}{{Agostinelli}, S.} \emph{et~al.}
\newblock \bibinfo{title}{{GEANT4{\textemdash}a simulation toolkit}}.
\newblock \emph{\bibinfo{journal}{Nucl. Instrum. Methods Phys. Res. Sect. A Accel. Spectrom. Detect. Assoc. Equip.}} \textbf{\bibinfo{volume}{506}}, \bibinfo{pages}{250--303}
  (\bibinfo{year}{2003}).

\bibitem{ATREYA2003}
\bibinfo{author}{{Atreya}, S.}, \bibinfo{author}{{Mahaffy}, P.},
  \bibinfo{author}{{Niemann}, H.}, \bibinfo{author}{{Wong}, M.} \&
  \bibinfo{author}{{Owen}, T.}
\newblock \bibinfo{title}{{Composition and origin of the atmosphere of Jupiter
  — an update, and implications for the extrasolar giant planets}}.
\newblock \emph{\bibinfo{journal}{Planet. Space Sci.}}
  \textbf{\bibinfo{volume}{51}}, \bibinfo{pages}{105 -- 112}
  (\bibinfo{year}{2003}).

\bibitem{Atreya1981}
\bibinfo{author}{{Atreya}, S.~K.}, \bibinfo{author}{{Donahue}, T.~M.} \&
  \bibinfo{author}{{Festou}, M.}
\newblock \bibinfo{title}{{Jupiter - Structure and composition of the upper
  atmosphere}}.
\newblock \emph{\bibinfo{journal}{\apjl}} \textbf{\bibinfo{volume}{247}},
  \bibinfo{pages}{L43--L47} (\bibinfo{year}{1981}).


\bibitem{Woodger2015}
\bibinfo{author}{{Woodger}, L.~A.} \emph{et~al.}
\newblock \bibinfo{title}{{A summary of the BARREL campaigns: Technique for
  studying electron precipitation}}.
\newblock \emph{\bibinfo{journal}{J. Geophys. Res. Space Phys.}} \textbf{\bibinfo{volume}{120}}, \bibinfo{pages}{4922--4935}
  (\bibinfo{year}{2015}).

\bibitem{Saur2006}
\bibinfo{author}{{Saur}, J.} \emph{et~al.}
\newblock \bibinfo{title}{{Anti-planetward auroral electron beams at Saturn}}.
\newblock \emph{\bibinfo{journal}{\nat}} \textbf{\bibinfo{volume}{439}},
  \bibinfo{pages}{699--702} (\bibinfo{year}{2006}).


%




















\bibitem{Bhardwaj2007}
\bibinfo{author}{{Bhardwaj}, A.} \emph{et~al.}
\newblock \bibinfo{title}{{X-rays from Solar System objects}}.
\newblock \emph{\bibinfo{journal}{Planet. Space Sci.}} \textbf{\bibinfo{volume}{55}},
  \bibinfo{pages}{1135--1189} (\bibinfo{year}{2007}).
\newblock \eprint{1012.1088}.

\bibitem{Dunn2020}
\bibinfo{author}{{Dunn}, W.} \emph{et~al.}
\newblock \bibinfo{title}{{In Search of X-rays from Uranus}}.
\newblock In \emph{\bibinfo{booktitle}{European Planetary Science Congress}},
  \bibinfo{pages}{EPSC2020--1028} (\bibinfo{year}{2020}).

\bibitem{Bhardwaj2006}
\bibinfo{author}{{Bhardwaj}, A.} \emph{et~al.}
\newblock \bibinfo{title}{{Low- to middle-latitude X-ray emission from
  Jupiter}}.
\newblock \emph{\bibinfo{journal}{J. Geophys. Res. Space Phys.}} \textbf{\bibinfo{volume}{111}}, \bibinfo{pages}{A11225}
  (\bibinfo{year}{2006}).

\bibitem{Imhof1974}
\bibinfo{author}{{Imhof}, W.~L.}, \bibinfo{author}{{Nakano}, G.~H.},
  \bibinfo{author}{{Johnson}, R.~G.} \& \bibinfo{author}{{Reagan}, J.~B.}
\newblock \bibinfo{title}{{Satellite observations of Bremsstrahlung from
  widespread energetic electron precipitation events}}.
\newblock \emph{\bibinfo{journal}{J. Geophys. Res.}} \textbf{\bibinfo{volume}{79}},
  \bibinfo{pages}{565--574} (\bibinfo{year}{1974}).

\bibitem{Chaston2008}
\bibinfo{author}{{Chaston}, C.~C.} \emph{et~al.}
\newblock \bibinfo{title}{{The Turbulent Alfv{\'e}nic Aurora}}.
\newblock \emph{\bibinfo{journal}{Phys. Rev. Lett.}} \textbf{\bibinfo{volume}{100}},
  \bibinfo{pages}{175003} (\bibinfo{year}{2008}).

\bibitem{Christensen2019}
\bibinfo{author}{{Christensen}, U.~R.}
\newblock \emph{\bibinfo{title}{{Planetary Magnetic Fields and Dynamos}}},
  \bibinfo{pages}{31} (\bibinfo{year}{2019}).
\newblock
  \urlprefix\url{https://doi.org/10.1093/acrefore/9780190647926.013.31}.

\bibitem{Bhardwaj2000}
\bibinfo{author}{{Bhardwaj}, A.} \& \bibinfo{author}{{Gladstone}, G.~R.}
\newblock \bibinfo{title}{{Auroral emissions of the giant planets}}.
\newblock \emph{\bibinfo{journal}{Rev. Geophys.}}
  \textbf{\bibinfo{volume}{38}}, \bibinfo{pages}{295--354}
  (\bibinfo{year}{2000}).

\bibitem{Feldman2000}
\bibinfo{author}{{Feldman}, P.~D.} \emph{et~al.}
\newblock \bibinfo{title}{{HST/STIS Ultraviolet Imaging of Polar Aurora on
  Ganymede}}.
\newblock \emph{\bibinfo{journal}{\apj}} \textbf{\bibinfo{volume}{535}},
  \bibinfo{pages}{1085--1090} (\bibinfo{year}{2000}).
\newblock \eprint{astro-ph/0003486}.

\bibitem{Delamere2004}
\bibinfo{author}{{Delamere}, P.~A.}, \bibinfo{author}{{Steffl}, A.} \&
  \bibinfo{author}{{Bagenal}, F.}
\newblock \bibinfo{title}{{Modeling temporal variability of plasma conditions
  in the Io torus during the Cassini era}}.
\newblock \emph{\bibinfo{journal}{J. Geophys. Res. Space Phys.}} \textbf{\bibinfo{volume}{109}}, \bibinfo{pages}{A10216}
  (\bibinfo{year}{2004}).

\bibitem{Dougherty2017}
\bibinfo{author}{{Dougherty}, L.~P.}, \bibinfo{author}{{Bodisch}, K.~M.} \&
  \bibinfo{author}{{Bagenal}, F.}
\newblock \bibinfo{title}{{Survey of Voyager plasma science ions at Jupiter: 2.
  Heavy ions}}.
\newblock \emph{\bibinfo{journal}{J. Geophys. Res. Space Phys.}} \textbf{\bibinfo{volume}{122}}, \bibinfo{pages}{8257--8276}
  (\bibinfo{year}{2017}).
  
\bibitem{Broadfoot1979}
\bibinfo{author}{{Broadfoot}, A.~L.} \emph{et~al.}
\newblock \bibinfo{title}{{Extreme Ultraviolet Observations from Voyager 1
  Encounter with Jupiter}}.
\newblock \emph{\bibinfo{journal}{Science}} \textbf{\bibinfo{volume}{204}},
  \bibinfo{pages}{979--982} (\bibinfo{year}{1979}).

\bibitem{Metzger1983}
\bibinfo{author}{{Metzger}, A.~E.} \emph{et~al.}
\newblock \bibinfo{title}{{The detection of X rays from Jupiter}}.
\newblock \emph{\bibinfo{journal}{J. Geophys. Res.}} \textbf{\bibinfo{volume}{88}},
  \bibinfo{pages}{7731--7741} (\bibinfo{year}{1983}).

\bibitem{Gerard2019}
\bibinfo{author}{{G{\'e}rard}, J.~C.} \emph{et~al.}
\newblock \bibinfo{title}{{Contemporaneous Observations of Jovian Energetic
  Auroral Electrons and Ultraviolet Emissions by the Juno Spacecraft}}.
\newblock \emph{\bibinfo{journal}{J. Geophys. Res. Space Phys.}} \textbf{\bibinfo{volume}{124}}, \bibinfo{pages}{8298--8317}
  (\bibinfo{year}{2019}).

\bibitem{Gladstone2002}
\bibinfo{author}{{Gladstone}, G.~R.} \emph{et~al.}
\newblock \bibinfo{title}{{A pulsating auroral X-ray hot spot on Jupiter}}.
\newblock \emph{\bibinfo{journal}{\nat}} \textbf{\bibinfo{volume}{415}},
  \bibinfo{pages}{1000--1003} (\bibinfo{year}{2002}).
  
\bibitem{BR2004}
\bibinfo{author}{{Branduardi-Raymont}, G.} \emph{et~al.}
\newblock \bibinfo{title}{{First observation of Jupiter by XMM-Newton}}.
\newblock \emph{\bibinfo{journal}{\aap}} \textbf{\bibinfo{volume}{424}},
  \bibinfo{pages}{331--337} (\bibinfo{year}{2004}).
\newblock \eprint{astro-ph/0406340}.


\bibitem{Paranicas2018}
\bibinfo{author}{{Paranicas}, C.} \emph{et~al.}
\newblock \bibinfo{title}{{Intervals of Intense Energetic Electron Beams Over
  Jupiter's Poles}}.
\newblock \emph{\bibinfo{journal}{J. Geophys. Res. Space Phys.}} \textbf{\bibinfo{volume}{123}}, \bibinfo{pages}{1989--1999}
  (\bibinfo{year}{2018}).
  
\bibitem{Kollmann2018}
\bibinfo{author}{{Kollmann}, P.} \emph{et~al.}
\newblock \bibinfo{title}{{Electron Acceleration to MeV Energies at Jupiter and
  Saturn}}.
\newblock \emph{\bibinfo{journal}{J. Geophys. Res. Space Phys.}} \textbf{\bibinfo{volume}{123}}, \bibinfo{pages}{9110--9129}
  (\bibinfo{year}{2018}).

\bibitem{Abazajian2009}
\bibinfo{author}{{Abazajian}, K.~N.} \emph{et~al.}
\newblock \bibinfo{title}{{The Seventh Data Release of the Sloan Digital Sky
  Survey}}.
\newblock \emph{\bibinfo{journal}{\apjs}} \textbf{\bibinfo{volume}{182}},
  \bibinfo{pages}{543--558} (\bibinfo{year}{2009}).
\newblock \eprint{0812.0649}.

\bibitem{Wik2014}
\bibinfo{author}{{Wik}, D.~R.} \emph{et~al.}
\newblock \bibinfo{title}{{NuSTAR Observations of the Bullet Cluster:
  Constraints on Inverse Compton Emission}}.
\newblock \emph{\bibinfo{journal}{\apj}} \textbf{\bibinfo{volume}{792}},
  \bibinfo{pages}{48} (\bibinfo{year}{2014}).
\newblock \eprint{1403.2722}.

\bibitem{Scargle2013}
\bibinfo{author}{{Scargle}, J.~D.}, \bibinfo{author}{{Norris}, J.~P.},
  \bibinfo{author}{{Jackson}, B.} \& \bibinfo{author}{{Chiang}, J.}
\newblock \bibinfo{title}{{Studies in Astronomical Time Series Analysis. VI.
  Bayesian Block Representations}}.
\newblock \emph{\bibinfo{journal}{\apj}} \textbf{\bibinfo{volume}{764}},
  \bibinfo{pages}{167} (\bibinfo{year}{2013}).
\newblock \eprint{1207.5578}.

\bibitem{JADE}
\bibinfo{author}{{McComas}, D.} \emph{et~al.}
\newblock \bibinfo{title}{{The Jovian Auroral Distributions Experiment (JADE)
  on the Juno Mission to Jupiter}}.
\newblock \emph{\bibinfo{journal}{Space Sci. Rev.}}
  \textbf{\bibinfo{volume}{213}}, \bibinfo{pages}{547--643}
  (\bibinfo{year}{2017}).

\bibitem{JEDI}
\bibinfo{author}{{Mauk}, B.} \emph{et~al.}
\newblock \bibinfo{title}{{The Jupiter Energetic Particle Detector Instrument
  (JEDI) Investigation for the Juno Mission}}.
\newblock \emph{\bibinfo{journal}{Space Sci. Rev.}}
  \textbf{\bibinfo{volume}{98}}, \bibinfo{pages}{98--} (\bibinfo{year}{2013}).

\bibitem{JADEDATA}
\bibinfo{author}{{Allegrini}, F.}, \bibinfo{author}{{Wilson}, R.},
  \bibinfo{author}{{Ebert}, R.} \& \bibinfo{author}{{Loeffler}, C.}
\newblock \bibinfo{title}{{JUNO JADE Calibrated Science Data}}
  (\bibinfo{publisher}{NASA Planetary Data System}, \bibinfo{year}{2019}).
\newblock
  \urlprefix\url{https://pds-ppi.igpp.ucla.edu/data/JNO-J_SW-JAD-3-CALIBRATED-V1.0/}.

\bibitem{JEDIDATA}
\bibinfo{title}{{JUNO JEDI Jupiter Standard Calibrated Products}}
  (\bibinfo{publisher}{NASA Planetary Data System}, \bibinfo{year}{2020}).
\newblock
  \urlprefix\url{https://pds-ppi.igpp.ucla.edu/data/JNO-J-JED-3-CDR-V1.0/}
  (\bibinfo{year}{2020}).

\bibitem{MAGDATA}
\bibinfo{author}{Connerney, J.}
\newblock \bibinfo{title}{{JUNO Magnetometer Jupiter Archive}}
  (\bibinfo{publisher}{NASA Planetary Data System}, \bibinfo{year}{2020}).
\newblock
  \urlprefix\url{https://pds-ppi.igpp.ucla.edu/data/JNO-J-3-FGM-CAL-V1.0/}
  (\bibinfo{year}{2020}).

\bibitem{SPICE}
\bibinfo{author}{{Acton}, C.}, \bibinfo{author}{{Bachman}, N.},
  \bibinfo{author}{{Semenov}, B.} \& \bibinfo{author}{{Wright}, E.}
\newblock \bibinfo{title}{{A look towards the future in the handling of space
  science mission geometry}}.
\newblock \emph{\bibinfo{journal}{Planet. Space Sci.}}
  \textbf{\bibinfo{volume}{150}}, \bibinfo{pages}{9 -- 12}
  (\bibinfo{year}{2018}).

\bibitem{Allegrini2020}
\bibinfo{author}{{Allegrini}, F.} \emph{et~al.}
\newblock \bibinfo{title}{{Energy Flux and Characteristic Energy of Electrons
  Over Jupiter's Main Auroral Emission}}.
\newblock \emph{\bibinfo{journal}{J. Geophys. Res. Space Phys.}} \textbf{\bibinfo{volume}{125}}, \bibinfo{pages}{e2019JA027693}.

\bibitem{AURORALOVALS}
\bibinfo{author}{Bonfond, B.} \emph{et~al.}
\newblock \bibinfo{title}{{Auroral evidence of Io's control over the
  magnetosphere of Jupiter}}.
\newblock \emph{\bibinfo{journal}{Geophys. Res. Lett.}}
  \textbf{\bibinfo{volume}{39}}.

\bibitem{Allison2006}
\bibinfo{author}{{Allison}, J.} \emph{et~al.}
\newblock \bibinfo{title}{{Geant4 developments and applications}}.
\newblock \emph{\bibinfo{journal}{IEEE Trans. Nucl. Sci.}}
  \textbf{\bibinfo{volume}{53}}, \bibinfo{pages}{270--278}
  (\bibinfo{year}{2006}).


\bibitem{PANDOLA2015}
\bibinfo{author}{Pandola, L.}, \bibinfo{author}{Andenna, C.} \&
  \bibinfo{author}{Caccia, B.}
\newblock \bibinfo{title}{{Validation of the GEANT4 simulation of
  bremsstrahlung from thick targets below 3 MeV}}.
\newblock \emph{\bibinfo{journal}{Nucl Instrum Methods Phys Res B}}
  \textbf{\bibinfo{volume}{350}}, \bibinfo{pages}{41 -- 48}
  (\bibinfo{year}{2015}).

\bibitem{Livengood1990}
\bibinfo{author}{Livengood, T.~A.}, \bibinfo{author}{Strobel, D.~F.} \&
  \bibinfo{author}{Moos, H.~W.}
\newblock \bibinfo{title}{{Long-term study of longitudinal dependence in
  primary particle precipitation in the north Jovian aurora}}.
\newblock \emph{\bibinfo{journal}{J. Geophys. Res. Space Phys.}} \textbf{\bibinfo{volume}{95}}, \bibinfo{pages}{10375--10388}.

\bibitem{Valek2019}
\bibinfo{author}{Valek, P.~W.} \emph{et~al.}
\newblock \bibinfo{title}{{Jovian High-Latitude Ionospheric Ions: Juno In Situ
  Observations}}.
\newblock \emph{\bibinfo{journal}{Geophys. Res. Lett.}}
  \textbf{\bibinfo{volume}{46}}, \bibinfo{pages}{8663--8670}.

\bibitem{Ozak2010}
\bibinfo{author}{{Ozak}, N.}, \bibinfo{author}{{Schultz}, D.~R.},
  \bibinfo{author}{{Cravens}, T.~E.}, \bibinfo{author}{{Kharchenko}, V.} \&
  \bibinfo{author}{{Hui}, Y.~W.}
\newblock \bibinfo{title}{{Auroral X-ray emission at Jupiter: Depth effects}}.
\newblock \emph{\bibinfo{journal}{J. Geophys. Res. Space Phys.}} \textbf{\bibinfo{volume}{115}}, \bibinfo{pages}{A11306}
  (\bibinfo{year}{2010}).

\bibitem{BR2008}
\bibinfo{author}{{Branduardi-Raymont}, G.} \emph{et~al.}
\newblock \bibinfo{title}{{Spectral morphology of the X-ray emission from
  Jupiter's aurorae}}.
\newblock \emph{\bibinfo{journal}{J. Geophys. Res. Space Phys.}} \textbf{\bibinfo{volume}{113}}, \bibinfo{pages}{A02202}
  (\bibinfo{year}{2008}).




\end{thebibliography}


\end{document}